\documentclass[reprint,amsmath,amssymb,aps,superscriptaddress]{revtex4-2}
\usepackage[english]{babel}
\makeatletter
\let\oldselectlanguage\selectlanguage
\renewcommand{\selectlanguage}[1]{%
  \def\tempa{#1}%
  \def\tempb{en}%
  \ifx\tempa\tempb
    \oldselectlanguage{english}%
  \else
    \oldselectlanguage{#1}%
  \fi
}
\makeatother

\usepackage[utf8]{inputenc}
\usepackage{adjustbox}
\usepackage{afterpage}
\usepackage{siunitx}
\usepackage{float}
\usepackage{upgreek}
\usepackage[dvipsnames]{xcolor}
\usepackage[normalem]{ulem}

\usepackage{hyperref}

\begin{document}

\title{Near-surface colloidal dynamics in jammed and slipping microgel suspensions
}

\author{Masoodah Gunny}
\affiliation{Gulliver CNRS UMR 7083, PSL Research University, ESPCI Paris, 10 rue Vauquelin, 75005 Paris, France}
\affiliation{IPGG, 6 rue Jean Calvin, 75005 Paris, France}
\author{Frédérick Caetano}
\affiliation{Université Lyon 1, CNRS, Institut Lumière Matière, UMR5306, F-69100, Villeurbanne, France}
\author{Matilde Bureau}
\affiliation{Université Lyon 1, CNRS, Institut Lumière Matière, UMR5306, F-69100, Villeurbanne, France}
\author{Alexandre Vilquin}
\affiliation{Gulliver CNRS UMR 7083, PSL Research University, ESPCI Paris, 10 rue Vauquelin, 75005 Paris, France}
\affiliation{IPGG, 6 rue Jean Calvin, 75005 Paris, France}
\author{Marie Le Merrer}
\email{marie.le-merrer@univ-lyon1.fr}
\affiliation{Université Lyon 1, CNRS, Institut Lumière Matière, UMR5306, F-69100, Villeurbanne, France}
\author{Joshua D. McGraw}
\email{joshua.mcgraw@espci.fr}
\affiliation{Gulliver CNRS UMR 7083, PSL Research University, ESPCI Paris, 10 rue Vauquelin, 75005 Paris, France}
\affiliation{IPGG, 6 rue Jean Calvin, 75005 Paris, France}
\author{Catherine Barentin}
\email{catherine.barentin@univ-lyon1.fr}
\affiliation{Université Lyon 1, CNRS, Institut Lumière Matière, UMR5306, F-69100, Villeurbanne, France}

\date{\today} 


\begin{abstract}
Jammed suspensions of soft microgel particles may exhibit slippage along smooth boundaries. Owing to their expected sub-micrometric dimensions, direct observations of dynamics within the near-surface layers supposed to be responsible for this slippage have been difficult to achieve. Here, we use total internal reflection fluorescence microscopy (TIRFM) to observe nanoparticle dynamics near glass/microgel-suspension interfaces. Indicating near-wall dynamic heterogeneity, velocity profiles for suspensions are nonlinear. These profiles tend to a constant slippage velocity at sub-micrometric distances from the wall, consistent with macroscopic wall slip measurements. Furthermore, nanoscale particle altitude distributions are strongly dependent on the slip velocity, revealing a dynamically-mediated and nanoscale particle-organisation effect. The collected observations give support for the existence of near-wall heterogeneity as a dominant mechanism contributing to microgel wall slip. Our work also opens new perspectives for the study of particle dynamics and organisation in complex interfacial environments. 

\end{abstract}


\maketitle

When a fluid in motion meets a solid, it may slide along the surface. For molecular fluids, important slip effects are generally confined to nanoscale systems not much larger than the constituent molecules~\cite{Neto2005, Joly2006, Cottin-Bizonne2008, Bocquet2010}. In contrast, macromolecular materials can exhibit slip conditions that are felt at distances orders of magnitude larger than the molecular components~\cite{malkin_wall_2018}. This contrast results mainly from the strong macromolecular size dependence of bulk dynamic quantities such as the viscosity, while surface friction can remain relatively low~\cite{degennes1979}. 
Polymer melts and solutions~\cite{degennes1979, Brochard-Wyart1996, Leger1997, Baumchen2009, Henot2018, Ilton2018, BarnesReview1995, Cuenca2013, Barraud2019, Cross2018, Park2019, Grzelka2021, Guyard2021}, emulsions~\cite{salmon_towards_2003, Goyon2008, Huerre2015, Mansard2014}, foams~\cite{denkov_wall_2005, marze_aqueous_2008, cantat_liquid_2013, le_merrer_linear_2015} and, as considered here, microgel suspensions~\cite{Meeker2004, Meeker2004PRL, Divoux2015, Jalaal2015, Zhang2017, Pemeja2019} can exhibit large slip effects along smooth walls. 

\begin{figure}[t!]
	\includegraphics[width=\columnwidth]{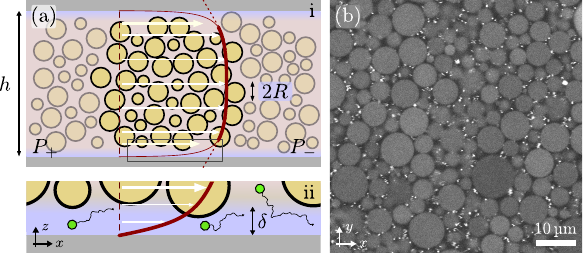}
    \caption{(a) Illustration of velocimetry experiments. (i) A channel with height $h$ is filled with a polydisperse, jammed suspension of microgel particles with average particle radius $R$. Pressure-driven flow induces a plug-like velocity profile (dark red, white arrows) with apparent slip (dotted lines). (ii) Zoom on a near-surface flow near a lubrication layer of size $\delta$. Fluorescent particles with radius $a$ are tracked using µPIV (i,  not shown) and TIRFM (ii, green). (b) Mid-channel confocal slice of stained microgels (Rhodamine 6G, large grey circles) and $a=55$~nm nanoparticles (small bright spots). }
  \label{fig:schematics}
\end{figure}

Biphasic systems~\cite{Coussot2007, Bonn2017, Cloitre2017} are often assumed to exhibit apparent slip. In this scenario, heterogeneity in the vicinity of the solid-fluid interface gives a macroscopic response which is slip-like, accompanied by an apparent velocity discontinuity at the wall as depicted in Figure~\ref{fig:schematics}(a.i, dashed line). At the microscale, however, the existence of a solvent-enriched, non-slipping layer, as schematically indicated in Figure~\ref{fig:schematics}(a.ii), is typically assumed~\cite{Grzelka2021, Guyard2021, Meeker2004PRL, Divoux2015}. This near-surface region, the subject of this Letter, is known as a depletion~\cite{Park2019, Barraud2019, Guyard2021} or lubrication~\cite{Meeker2004PRL, Skotheim2004} layer. 

The size of supposed lubrication layers depends on the specific fluid. For polymer solutions~\cite{BarnesReview1995, Barraud2019, Cross2018, Park2019, Grzelka2021, Guyard2021}, the appropriate scale is the distance between chains in bulk solution~\cite{Rubinstein2003, Joanny1979, Dobrynin1995, Lafon2025} (\emph{i.e.}, the correlation length), typically a few to tens of nanometers. For confined droplets or bubbles, reminiscent of emulsions or foams, interferometric measurements revealed thicknesses in the range of tens to hundreds of nanometers~\cite{Huerre2015, denkov_foam_2006, Reichert2018}. Such length scales are consistent with predictions based on electrostatic or visco-capillary hydrodynamic interactions between the solid-fluid and fluid-fluid interfaces.

For microgel suspensions, similar considerations apply~\cite{seth_influence_2008, Meeker2004PRL, Snoeijer2013}. In particular, an elastohydrodynamic (EHD) balance~\cite{Skotheim2004} at the wall can be used to predict lubrication layer thicknesses, $\delta$, of order tens to hundreds of nanometers and increasing with the slip velocity, $V_\mathrm{s}$~\cite{Meeker2004PRL, Snoeijer2013}. However, poor optical contrast in solvent-microgel suspensions ---typically less than 1\% polymer--- has limited the study of microgel lubrication layers to their macroscopic signatures. Expressed in terms of a friction law, these signatures link the wall stress and slip velocity. With $\eta_0$ the near-wall viscosity, the wall stress is viscous-like, $\sigma_\mathrm{w} \sim \eta_0 V_\mathrm{s}/\delta$, and, since $\delta$ may depend on the slip velocity, nonlinear friction laws are common. Validation of nonlinear, macroscopic friction laws~\cite{Meeker2004PRL, geraud_confined_2013, Divoux2015, ortega-avila_axial_2016, Zhang2017, Pemeja2019} provided indirect evidence for near-wall dynamic heterogeneity in microgel suspensions. 

While these aforementioned laws assume smooth gel-interface profiles, at the microscale, the near-wall region is structurally heterogeneous. This region contains rough, interstitial spaces between gel particles, and, as for a majority of real, jammed systems, a high variability of microgel sizes as seen in Figure~\ref{fig:schematics}. Under motion, this structural heterogeneity may thus generate important structural and dynamical consequences that we aim to elucidate here. We use total internal reflection fluorescence microscopy (TIRFM, based on nanoparticle tracking) complemented by microparticle image velocimetry (µPIV) to investigate this complex space. First, a fluidised interfacial region much smaller than the microgel size is revealed, the dynamics therein rationalised by considering the microscopic roughness. Second, a dynamically-mediated and near-surface nanoparticle enrichment  is observed, assumed to be piloted by diffential EHD interactions~\cite{Skotheim2004, Meeker2004PRL, McGraw2025}, between microgels and the former nanoparticles.

\begin{figure}[b!]
    \centering
    \includegraphics[scale=1, trim=0 1mm 0 0, clip]{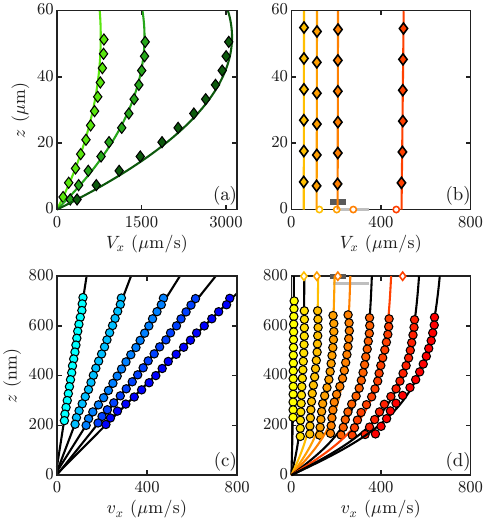}
   \caption{Velocity profiles obtained from (a),(b) µPIV and (c),(d) TIRFM for (a),(c) Newtonian liquids (75wt\% glycerol-water mixture and ultra-pure water, respectively) and (b),(d) microgel suspensions. The lines represent fits to (a) Poiseuille, (b) plug, (c) linear-shear, and (d) nonlinear profiles. Applied pressures follow a common color code in (b),(d), from 100~mbar to 700~mbar. Correspondence between the two techniques are indicated by the circles and lozenges added on the bottom and top axes of (b) and (d). Error bars for 500 mbar data sets, representative for all pressures, are indicated for µPIV (dark grey) and TIRFM (light grey). Pressures and fitting parameters are given in Table S1 (S2.1).  }
    \label{fig:velocimetry}
\end{figure}

We used aqueous suspensions of microgels composed of crosslinked sodium polyacrylate {(FLOPRINT TA 150 A from SNF)}. When dispersed in water at concentration 2~g/L (see Supplementary Material S1.1 in Ref.~\footnote{%
SM contains details on: experimental protocols; fitting velocity profiles of Figure~\ref{fig:velocimetry}; uncertainty on the velocity and comparison between TIRFM and µPIV velocites; information supporting enhanced-viscosity lubrication layer; influence of solvent-microgel boundary shape on measured velocity; derivation for the elastohydrodynamic balance to see nanoparticles at the wall%
} %
for the detailed preparation protocol), the polymer microgels swell and form jammed spheres with characteristic radius $2 \lesssim R\lesssim 3$~$\mu$m (S1.2), as seen in Figure~\ref{fig:schematics}: schematically in (a) and using confocal microscopy in (b). The suspension is a yield stress fluid~\cite{Bonn2017} (see S1.3 for rheology). Fluorescent particles of radius $a$ are incorporated in the suspension for flow visualisation, with $a = 500$~nm for µPIV and $a = 55$~nm for TIRFM (Invitrogen F8820 and F8803, the latter termed nanoparticles), both at volume fractions $\phi \approx 10^{-5}$.

We observed microgel suspension flows using µPIV and TIRFM in microchannels with dimensions $\{L,w,h\} = \{40\,\mathrm{mm}, 2\,\mathrm{mm},110\,\mu\mathrm{m}\}$\footnote{The reported height corresponds to the average channel height under flow, as measured by confocal microscopy} along the $\{x,y,z\}$ directions. An applied pressure difference, $P$ (Elveflow controllers) drove flows along the $x$ direction. Channels were fabricated from PDMS (Momentive RTV, 10 wt\%  cross linker, cured at 70~$^\circ$C for 3 h) before plasma bonding to glass cover slips. Plastic tubing (Tygon ND 100-80, 0.57\,mm ID, 60~cm long) connected the channels to liquid reservoirs. Suspensions were pre-conditioned~\cite{divoux_transient_2010} with an excess pressure between each working pressure (S1.4). TIRFM velocimetry results were linked to µPIV wall-stress measurements using glass capillaries with $\{L,w,h\} = \{50\,\mathrm{mm},3\,\mathrm{mm}, 300\,\mu\mathrm{m}\}$ directly connected to fluid reservoirs and with identical suspensions~\footnote{For TIRFM measurements the wall position must be determined \emph{a priori} with a calibrating water measurement as in Ref.~\cite{Guyard2021}; yet, the necessity of two liquid reservoirs requires intermediate tubings between reservoir and chip.}. 

Flows profiles across the whole channel were accessed using µPIV as in Refs.~\cite{geraud_confined_2013, Pemeja2019}. Briefly, a fluorescence microscope was used to image the positions of nanoparticles at different heights, $z$, above the solid-liquid interface. The vertical resolution of a few microns is limited by the field depth of the objective (Leica HC PL Fluotar 20$\times$/0.40). Images in each $z$-plane were analyzed with either image correlation or particle tracking to determine the horizontal velocity, denoted $V_x(z)$ for vertical position $1\text{~µm} <z<h/2$. 
For near-surface flows, we used objective-based (Nikon APO TIRF, 100$\times$/1.49) TIRFM to track nanoparticles in 3D. This technique exploits the exponential decay of the evanescent field to obtain particle height based on fluorescence intensity, and particle tracking to obtain velocity~\cite{Huang2006, Yoda2011, Li2015, Guyard2021, Vilquin2021, Vilquin2023}. Velocimetry measurements at this scale, denoted $v_x(z)$, were first conducted in water at various pressures, serving as calibration experiments to determine the wall position as in Ref.~\cite{Guyard2021}. The channel was then flushed with ultrapure water until no fluorescent nanoparticles could be observed. Subsequently, nanoparticle-laden microgel suspensions were injected into the channels. Identical trends in all that follows were also observed in a more concentrated suspension at 2.4 g/L.

In Figure~\ref{fig:velocimetry}, we show velocity profiles measured with Newtonian fluids (left panels) and microgel suspensions (right panels) at two different scales. At the channel scale (top), the pressure driven flow of the Newtonian fluid shows a Poiseuille flow as expected; in contrast, the microgel suspension exhibits plug flow for all investigated pressures. At the microscopic scale (bottom), velocity profiles of the Newtonian fluid are linear. In contrast, and as shown in Figure~\ref{fig:velocimetry}(d), velocity profiles for the microgel suspensions are non-linear. A highly sheared region close to the wall coexists with a non-sheared region beyond it; assuming continuity of shear stress, the coexistence between the highly-sheared and non-sheared regions is a signature for the coexistence between regions of low and high viscosity. This observation gives a first support for the existence of a solvent-rich lubrication layer near the wall, as sketched in Figure~\ref{fig:schematics}(a.ii). 

\begin{figure}[b!]
	\includegraphics{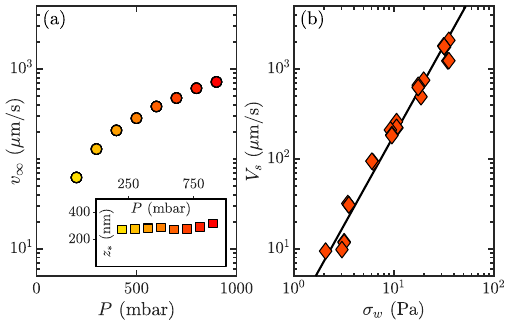}
    \caption{(a) Far-field velocity $v_\infty$ as a function of the applied pressure $P$ in the PDMS-glass channels. Inset: characteristic length $z_*$ as a function of $P$. Fit parameters extracted from data in Figure~\ref{fig:velocimetry}(d). (b) Friction law measured using µPIV in rectangular glass capillaries for the microgel suspension: slip velocity $V_\mathrm{s}$ as a function of the wall shear stress $\sigma_\mathrm{w}$; the black line has log-slope 2 as described in the text. 
    }
  \label{fig:friction_laws}
\end{figure}

Nonlinear velocity profiles in Figure~\ref{fig:velocimetry}(d) were fit using the empirical relation $v_x(z)=v_\infty(1+(z_*/z)^k)^{-1/k}$. The fitting is summarised in Figure~\ref{fig:friction_laws}(a): $v_\infty$ is the large-$z$ velocity asymptote and $z_*$ (inset) characterises the relatively constant position of the slope rupture of order a few hundred nanometers. We fix $k=2$~\footnote{
The exponent $k$ mediates the transition between linear and constant velocity regimes. Choosing other values of $k$ has no impact on general conclusions reached here but contributes to the error bars on $z_*$%
}. %
The values of $v_\infty$ ---consistent with the slip velocities measured using µPIV as shown in Figures~\ref{fig:velocimetry}(b) and (d) and in S2.2--- increase with the applied pressure as seen in Figure~\ref{fig:friction_laws}(a). Connecting $v_\infty$ to the wall stress is done using rectangular glass capillaries, for which we have $\sigma_\mathrm{w} = Ph/2L$ following Ref.~\cite{Pemeja2019}. We thus obtain, from µPIV, the friction law shown in Figure~\ref{fig:friction_laws}(b). The solid line passing through the data is given by the relation $V_\mathrm{s} = \alpha \sigma_\mathrm{w}^2$, with $\alpha = 1.8\pm0.2$\,µm\,s$^{-1}$\,Pa$^{-2}$, and is consistent with previous measurements~\cite{Pemeja2019}. We now discuss these observations in terms of wall slippage literature and interfacial roughness hypotheses described above. 

First, we note that the reported $z_* \approx 10^2$\,nm is one order of magnitude smaller than the microgel radius, $R$. This observation shows that the effective fluidisation at the wall cannot be attributed to a reorganisation of the first microgel layers \cite{mansard_boundary_2014, younes_slippery_2020,  jung_wall_2021}. The pressure independence of $z_*$ suggests furthermore that if it were a length scale corresponding only to the lubrication layer thickness, $\delta$, a linear friction law would be expected. Non-linearity as observed in Figure~\ref{fig:friction_laws}(b), however, typically results from $\sigma_\mathrm{w} = \eta_0 V_\mathrm{s}/\delta(V_\mathrm{s})$, where $\delta$ is expected to increase with $V_s$ and to depend on fluid properties such as the near-wall viscosity $\eta_0$, which we now seek to estimate in order to build an accord between our micro- and macroscopic measurements.

Comparing Figures~\ref{fig:friction_laws}(a) and (b), we find that for $60\lesssim v_\infty \lesssim 700$\,µm\,s$^{-1}$, wall stresses in the range from 5 to 20~Pa are expected. Using $v_\infty/z_*\approx\dot{\gamma}_0$, and a viscous stress at the wall, $\sigma_\mathrm{w}=\eta_0\dot{\gamma}_0$, we have $\eta_0 \approx10$\,--\,20\,mPa\,s. Consistent with our control measurements in S2.2 for which nanoparticle diffusion and differing degrees of filtration of the microgel suspensions were investigated, this near-wall viscosity is larger than that of the solvent (1~mPa\,s). Such a large value could be explained by the presence of free polymer chains or submicrometric gel particles~\cite{agnolin_internal_2007, galvani_hierarchical_2023, McGraw2025} remaining near the wall under stress. Importantly, the near-wall viscosity estimate is between one and two orders of magnitude \emph{smaller} than the bulk viscosity of the microgel suspension at similar stress, \emph{cf.} S1.3. Such a viscosity estimate recalls our earlier observation that the nonlinear velocity profiles provide evidence for a heterogeneous near-surface region of relatively \emph{low} viscosity. The nontrivial boundary shape of this region, and its impact on ensemble-averaged nanoscale velocimetry, is discussed next. 

As indicated in Figure~\ref{fig:schematics}(a.ii), the gel-solvent boundary close to the wall is not flat, but should be corrugated due to the rounded shape of the microgels. Given the typical microgel size of a few microns as in Figure~\ref{fig:schematics}(b), we expect this corrugation to be of order 1~µm. Under this assumption, the typical height $z_*$ should correspond to an average between the micron-scale roughness, of order $R$, and $\delta \sim R(\eta_0 v_\infty/R G_\mathrm{p})^{1/2}\approx 30$\,--\,$160$~nm predicted by \cite{Meeker2004}\footnote{In Ref.~\cite{Snoeijer2013} is provided a different prediction $\delta \sim R(\eta_0 v_\infty/R G_\mathrm{p})^{3/5}$, which yields $\delta \approx 10-90$~nm.}. This latter estimate of $\delta$ is made using $\eta_\mathrm{0}$ quoted above, $G_\mathrm{p}\approx 1600$~Pa the microgel elastic modulus (see Figure~S1(b) in S1.3 and Ref.~\cite{Seth2006}) and $R\approx3$\,µm. Besides, the diameter of the nanoparticles used in TIRFM is $110$~nm, much larger than mesh sizes of a few to a few 10's of nanometers for typical gels (\emph{cf.} S1.3.2). The velocity profiles of Figure~\ref{fig:velocimetry}(d) are thus expected to report the average of that of the near-wall gel interstices only. 

In order to justify the observation that $z_*$ does not vary as much as the expected lubrication layer thicknesses, we make two simplifying hypotheses: (i) the roughness profile is periodic and 1D at microscales, $h_\upmu=f(x) = \delta + \varepsilon[1+\cos(\pi x/R)]$; and (ii) the flow is pure shear under the corrugated roughness, $v_x(z)|_{x=x_0}=v_\infty z/f(x_0)$. While the pure-shear hypothesis does not verify Navier-Stokes equation, it allows us to capture the impact of the microscopic roughness profile; see S3.1.2 for more details and perspectives. Spatially-averaged velocity profiles observed with TIRFM can then be approximated as $v_x(z) = v_\infty L^{-1}(z)\int_{\{x: f\geq z\}}zf^{-1}(x)\,\mathrm{d}x$. Here, $L(z)$ is the length of the region satisfying the integral bound, the latter which must contain solvent only. In S3.1.1, we show that the mean velocity profiles computed using this simplified model are indeed not linear. Consistent with the experimental data, the predicted profiles start with a large slope near the wall, and give smaller slopes as the upper limit of the roughness is reached at $z = \delta + 2\varepsilon$. For the range of $\delta$ presented above, and choosing identical $\varepsilon, R$, we find that a five-fold change in $\delta$ results in only a 20\% change in $z_*$. We conclude that corrugation models thus account in large part for the weak variation of $z_*$ with $P$ or $v_\infty$ observed here, and are a necessary consideration for these complex interfacial scenarios.

\begin{figure}[b!]
	\includegraphics{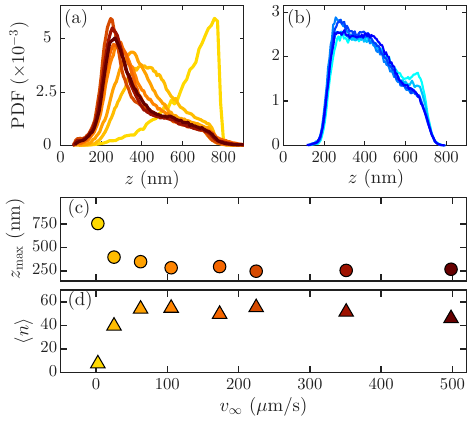}
    \caption{(a) Altitude probability distributions for 55 nm nanoparticles dispersed in {the} microgel suspension. Pressure increases from 50 mbar to 500 mbar from yellow to red. (b) Same measurements performed in pure water for different pressures. (c) Maximum probability altitude as a function of $v_\infty$. (d) Average number of particles in the TIRFM field of view, as a function of $v_\infty$ for the same data set as in (c).}
  \label{fig:SIDsx} 
\end{figure}

To complete the dynamical quantification and description of the rough geometry of the near-surface heterogeneous region, we now study how nanoparticles are organized there. Altitude probability distributions (\cite{Li2015,Zheng2018,Vilquin2021,Vilquin2023}, APDs) of nanoparticles in TIRFM are thus shown in Figure~\ref{fig:SIDsx}(a) for different pressures in microgel suspension. In contrast to the APDs for the same nanoparticles in water shown in the Figure~\ref{fig:SIDsx}(b), these former nanoparticle APDs in microgel suspensions are strongly and remarkably pressure dependent. For microgel suspensions, higher near-surface velocity is thus correlated to a nanoparticle enrichment near the glass-fluid interface. 

To quantify the nanoparticle enrichment, we show in Figure~\ref{fig:SIDsx} and as functions of the slipping velocity: first, the position of the APD maximum{, denoted $z_\text{max}$} (c); and second, the average number of particles in the field of view per frame, $\langle n\rangle$ (d). Consistent with the development of a layer partially depleted of microgel particles near the interface, we observe simultaneously that, as the slipping velocity increases: the {nano}particles observed are seen closer to the wall; and, more {nano}particles are observed in the evanescent zone. 

To explain these observations qualitatively, we note that since the nanoparticles used here are rigid, they are not subject to the EHD lift~\cite{Skotheim2004} that is exerted on soft microgels. Hypothetically, therefore~\cite{McGraw2025}, space liberated  through the EHD interaction on the microgels~\cite{Meeker2004} offers an entropic gain to the rigid nanoparticles, and they diffuse there. Given the typical sizes for the lubrication layers discussed above, however, $\delta \lesssim 2a$, the migration scenario at play here may implicate a deformation of the microgels. However, the growth of a lubrication layer even if smaller than the nanoparticle diameter, may nevertheless enhance the probability that a nanoparticle becomes transiently pinned at the surface. 

The energy required to indent an elastic microgel with modulus $G_\mathrm{p}$ by an amount $2a-\delta$ is roughly $U_\mathrm{el}\approx G_\mathrm{p}a^{1/2}(2a-\delta)^{5/2}$. This energy could be compared with the viscous dissipation, $Q\sim\eta_0\left(v_\infty/\delta\right)^2$, in the alternative process for which the nanoparticle at the wall goes around the microgel without indenting it. In S3.2, we estimate $Q\approx G_\mathrm{p}a^3((2a-\delta)R^{-1})^{1/2}$, where the modulus appears thanks to an invocation of Meeker's EHD microgel lubrication model~\cite{Meeker2004}. When dissipation is larger than the deformation energy, the latter is preferred, nanoparticles tend to go under the microgel and consequently an enrichment at the wall can be expected. 

The onset of the enrichment occurs when $U_\mathrm{el} = Q$, giving a critical lubrication-layer thickness that is strictly less than $a$. This critical velocity scales as $v_\mathrm{c}\approx G_\mathrm{p}a^2/\eta_0 R$, beyond which nanoparticles are expected to be enriched near the glass surface. Considering the typical values for $G_\mathrm{p}$, $a$, and $R$, and using $\eta_0 \approx 10$~mPa\,s as discussed above, we find a typical cross-over velocity of order $10^2$\,µm s$^{-1}$, in good agreement with our experimental observations on $z_\mathrm{max}$ and $\langle n\rangle$ in Figures~\ref{fig:SIDsx}(c) and (d), where nanoparticles are enriched near the wall for $v_\infty\gtrsim 100$\ µm s$^{-1}$. The existence of a critical velocity for near-surface particle enrichment provides another evidence for the existence of a lubrication layer. 

To conclude, our results are based on two particular features of the 3D, nanometric spatial resolution of TIRFM. First, for near-surface velocimetry, we observe a nonlinear velocity profile tending to a constant value. These nanoscale profiles are consistent with the apparent plug-flow at observed with µPIV at larger scale, but are also consistent with a no-slip condition at nanoscales. Second, the APDs display a marked dependence with the applied pressure, indicating a velocity-dependent interaction between the particles and the wall. This latter interaction is consistent with a liberation of space by the microgels during the creation of a lubrication layer. Collectively, the measurements constitute a first direct evidence of a lubrication layer for microgel wall slip. A quantitative comparison of our microscopic and macroscopic measurements suggests a key role of the microgel-induced roughness in the interfacial layer. These latter observations combined with the APDs thus open new perspectives for studying the near-surface dynamics in complex, heterogeneous media. 

\textbf{Acknowledgments} The authors gratefully acknowledge: Sébastien Manneville, Michel Cloitre, Frédéric Restagno and Nicolas Bain for enlightening discussions; Gilles Simon for technical support; and SNF for providing microgel samples. The authors acknowledge the Institut Pierre-Gilles de Gennes under PSL Research University's Major Research Program of the same name under ANR-10-IDEX-0001; they thank the joint service unit CNRS UAR 3750 for their continued technical support. The authors benefited from the financial support of the CoPinS ANR-19CE06-0021 grant. The work was also funded by the European Union under the NoDiCE ERC-2024-COG-101170653 grant. Views and opinions expressed are however those of the author(s) only and do not necessarily reflect those of the European Union or the European Research Council Executive Agency. Neither the European Union nor the granting authority can be held responsible for them. MG acknowledges Campus France for financial support. The authors also thank CNRS GDRs ISM and SLAMM along with the ICAM SliM-EX program for travel support.

\bibliography{../PAPER-uGelSlip.bib}

\begin{thebibliography}{65}%
\makeatletter
\providecommand \@ifxundefined [1]{%
 \@ifx{#1\undefined}
}%
\providecommand \@ifnum [1]{%
 \ifnum #1\expandafter \@firstoftwo
 \else \expandafter \@secondoftwo
 \fi
}%
\providecommand \@ifx [1]{%
 \ifx #1\expandafter \@firstoftwo
 \else \expandafter \@secondoftwo
 \fi
}%
\providecommand \natexlab [1]{#1}%
\providecommand \enquote  [1]{``#1''}%
\providecommand \bibnamefont  [1]{#1}%
\providecommand \bibfnamefont [1]{#1}%
\providecommand \citenamefont [1]{#1}%
\providecommand \href@noop [0]{\@secondoftwo}%
\providecommand \href [0]{\begingroup \@sanitize@url \@href}%
\providecommand \@href[1]{\@@startlink{#1}\@@href}%
\providecommand \@@href[1]{\endgroup#1\@@endlink}%
\providecommand \@sanitize@url [0]{\catcode `\\12\catcode `\$12\catcode
  `\&12\catcode `\#12\catcode `\^12\catcode `\_12\catcode `\%12\relax}%
\providecommand \@@startlink[1]{}%
\providecommand \@@endlink[0]{}%
\providecommand \url  [0]{\begingroup\@sanitize@url \@url }%
\providecommand \@url [1]{\endgroup\@href {#1}{\urlprefix }}%
\providecommand \urlprefix  [0]{URL }%
\providecommand \Eprint [0]{\href }%
\providecommand \doibase [0]{https://doi.org/}%
\providecommand \selectlanguage [0]{\@gobble}%
\providecommand \bibinfo  [0]{\@secondoftwo}%
\providecommand \bibfield  [0]{\@secondoftwo}%
\providecommand \translation [1]{[#1]}%
\providecommand \BibitemOpen [0]{}%
\providecommand \bibitemStop [0]{}%
\providecommand \bibitemNoStop [0]{.\EOS\space}%
\providecommand \EOS [0]{\spacefactor3000\relax}%
\providecommand \BibitemShut  [1]{\csname bibitem#1\endcsname}%
\let\auto@bib@innerbib\@empty
\bibitem [{\citenamefont {Neto}\ \emph {et~al.}(2005)\citenamefont {Neto},
  \citenamefont {Evans}, \citenamefont {Bonaccurso}, \citenamefont {Butt},\
  and\ \citenamefont {Craig}}]{Neto2005}%
  \BibitemOpen
  \bibfield  {author} {\bibinfo {author} {\bibfnamefont {C.}~\bibnamefont
  {Neto}}, \bibinfo {author} {\bibfnamefont {D.~R.}\ \bibnamefont {Evans}},
  \bibinfo {author} {\bibfnamefont {E.}~\bibnamefont {Bonaccurso}}, \bibinfo
  {author} {\bibfnamefont {H.-J.}\ \bibnamefont {Butt}},\ and\ \bibinfo
  {author} {\bibfnamefont {V.~S.~J.}\ \bibnamefont {Craig}},\ }\bibfield
  {title} {\bibinfo {title} {Boundary slip in {Newtonian} liquids: a review of
  experimental studies},\ }\href {https://doi.org/10.1088/0034-4885/68/12/R05}
  {\bibfield  {journal} {\bibinfo  {journal} {Reports on Progress in Physics}\
  }\textbf {\bibinfo {volume} {68}},\ \bibinfo {pages} {2859} (\bibinfo {year}
  {2005})}\BibitemShut {NoStop}%
\bibitem [{\citenamefont {Joly}\ \emph {et~al.}(2006)\citenamefont {Joly},
  \citenamefont {Ybert},\ and\ \citenamefont {Bocquet}}]{Joly2006}%
  \BibitemOpen
  \bibfield  {author} {\bibinfo {author} {\bibfnamefont {L.}~\bibnamefont
  {Joly}}, \bibinfo {author} {\bibfnamefont {C.}~\bibnamefont {Ybert}},\ and\
  \bibinfo {author} {\bibfnamefont {L.}~\bibnamefont {Bocquet}},\ }\bibfield
  {title} {{\selectlanguage {english}\bibinfo {title} {Probing the
  {Nanohydrodynamics} at {Liquid}-{Solid} {Interfaces} {Using} {Thermal}
  {Motion}}},\ }\href@noop {} {\bibfield  {journal} {\bibinfo  {journal}
  {Physical Review Letters}\ }\textbf {\bibinfo {volume} {96}},\ \bibinfo
  {pages} {046101} (\bibinfo {year} {2006})}\BibitemShut {NoStop}%
\bibitem [{\citenamefont {Cottin-Bizonne}\ \emph {et~al.}(2008)\citenamefont
  {Cottin-Bizonne}, \citenamefont {Steinberger}, \citenamefont {Cross},
  \citenamefont {Raccurt},\ and\ \citenamefont
  {Charlaix}}]{Cottin-Bizonne2008}%
  \BibitemOpen
  \bibfield  {author} {\bibinfo {author} {\bibfnamefont {C.}~\bibnamefont
  {Cottin-Bizonne}}, \bibinfo {author} {\bibfnamefont {A.}~\bibnamefont
  {Steinberger}}, \bibinfo {author} {\bibfnamefont {B.}~\bibnamefont {Cross}},
  \bibinfo {author} {\bibfnamefont {O.}~\bibnamefont {Raccurt}},\ and\ \bibinfo
  {author} {\bibfnamefont {E.}~\bibnamefont {Charlaix}},\ }\bibfield  {title}
  {{\selectlanguage {english}\bibinfo {title} {Nanohydrodynamics: {The}
  {Intrinsic} {Flow} {Boundary} {Condition} on {Smooth} {Surfaces}}},\ }\href
  {https://doi.org/10.1021/la7024044} {\bibfield  {journal} {\bibinfo
  {journal} {Langmuir}\ }\textbf {\bibinfo {volume} {24}},\ \bibinfo {pages}
  {1165} (\bibinfo {year} {2008})}\BibitemShut {NoStop}%
\bibitem [{\citenamefont {Bocquet}\ and\ \citenamefont
  {Charlaix}(2010)}]{Bocquet2010}%
  \BibitemOpen
  \bibfield  {author} {\bibinfo {author} {\bibfnamefont {L.}~\bibnamefont
  {Bocquet}}\ and\ \bibinfo {author} {\bibfnamefont {E.}~\bibnamefont
  {Charlaix}},\ }\bibfield  {title} {{\selectlanguage {english}\bibinfo {title}
  {Nanofluidics, from bulk to interfaces}},\ }\href
  {https://doi.org/10.1039/B909366B} {\bibfield  {journal} {\bibinfo  {journal}
  {Chem. Soc. Rev.}\ }\textbf {\bibinfo {volume} {39}},\ \bibinfo {pages}
  {1073} (\bibinfo {year} {2010})}\BibitemShut {NoStop}%
\bibitem [{\citenamefont {Malkin}\ and\ \citenamefont
  {Patlazhan}(2018)}]{malkin_wall_2018}%
  \BibitemOpen
  \bibfield  {author} {\bibinfo {author} {\bibfnamefont {A.~Y.}\ \bibnamefont
  {Malkin}}\ and\ \bibinfo {author} {\bibfnamefont {S.~A.}\ \bibnamefont
  {Patlazhan}},\ }\bibfield  {title} {\bibinfo {title} {Wall slip for complex
  liquids -- {Phenomenon} and its causes},\ }\href
  {https://doi.org/10.1016/j.cis.2018.05.008} {\bibfield  {journal} {\bibinfo
  {journal} {Advances in Colloid and Interface Science}\ }\textbf {\bibinfo
  {volume} {257}},\ \bibinfo {pages} {42} (\bibinfo {year} {2018})}\BibitemShut
  {NoStop}%
\bibitem [{\citenamefont {{de Gennes}}(1979)}]{degennes1979}%
  \BibitemOpen
  \bibfield  {author} {\bibinfo {author} {\bibfnamefont {P.-G.}\ \bibnamefont
  {{de Gennes}}},\ }\bibfield  {title} {\bibinfo {title} {{\'E}coulements
  viscom{\'e}triques de polym{\`e}res enchev{\^e}tr{\'e}s},\ }\href@noop {}
  {\bibfield  {journal} {\bibinfo  {journal} {C. R. Acad. Sci. Paris, S{\'e}rie
  B}\ }\textbf {\bibinfo {volume} {288}},\ \bibinfo {pages} {219} (\bibinfo
  {year} {1979})}\BibitemShut {NoStop}%
\bibitem [{\citenamefont {{Brochard-Wyart}}\ \emph {et~al.}(1996)\citenamefont
  {{Brochard-Wyart}}, \citenamefont {Gay},\ and\ \citenamefont {{de
  Gennes}}}]{Brochard-Wyart1996}%
  \BibitemOpen
  \bibfield  {author} {\bibinfo {author} {\bibfnamefont {F.}~\bibnamefont
  {{Brochard-Wyart}}}, \bibinfo {author} {\bibfnamefont {C.}~\bibnamefont
  {Gay}},\ and\ \bibinfo {author} {\bibfnamefont {P.-G.}\ \bibnamefont {{de
  Gennes}}},\ }\bibfield  {title} {\bibinfo {title} {Slippage of {{Polymer
  Melts}} on {{Grafted Surfaces}}},\ }\href@noop {} {\bibfield  {journal}
  {\bibinfo  {journal} {Macromolecules}\ }\textbf {\bibinfo {volume} {29}},\
  \bibinfo {pages} {377} (\bibinfo {year} {1996})}\BibitemShut {NoStop}%
\bibitem [{\citenamefont {L{\'e}ger}\ \emph {et~al.}(1997)\citenamefont
  {L{\'e}ger}, \citenamefont {Hervet}, \citenamefont {Massey},\ and\
  \citenamefont {Durliat}}]{Leger1997}%
  \BibitemOpen
  \bibfield  {author} {\bibinfo {author} {\bibfnamefont {L.}~\bibnamefont
  {L{\'e}ger}}, \bibinfo {author} {\bibfnamefont {H.}~\bibnamefont {Hervet}},
  \bibinfo {author} {\bibfnamefont {G.}~\bibnamefont {Massey}},\ and\ \bibinfo
  {author} {\bibfnamefont {E.}~\bibnamefont {Durliat}},\ }\bibfield  {title}
  {\bibinfo {title} {Wall slip in polymer melts},\ }\href
  {https://doi.org/10.1088/0953-8984/9/37/006} {\bibfield  {journal} {\bibinfo
  {journal} {Journal of Physics: Condensed Matter}\ }\textbf {\bibinfo {volume}
  {9}},\ \bibinfo {pages} {7719} (\bibinfo {year} {1997})}\BibitemShut
  {NoStop}%
\bibitem [{\citenamefont {B{\"a}umchen}\ \emph {et~al.}(2009)\citenamefont
  {B{\"a}umchen}, \citenamefont {Fetzer},\ and\ \citenamefont
  {Jacobs}}]{Baumchen2009}%
  \BibitemOpen
  \bibfield  {author} {\bibinfo {author} {\bibfnamefont {O.}~\bibnamefont
  {B{\"a}umchen}}, \bibinfo {author} {\bibfnamefont {R.}~\bibnamefont
  {Fetzer}},\ and\ \bibinfo {author} {\bibfnamefont {K.}~\bibnamefont
  {Jacobs}},\ }\bibfield  {title} {\bibinfo {title} {Reduced {{Interfacial
  Entanglement Density Affects}} the {{Boundary Conditions}} of {{Polymer
  Flow}}},\ }\href {https://doi.org/10.1103/PhysRevLett.103.247801} {\bibfield
  {journal} {\bibinfo  {journal} {Physical Review Letters}\ }\textbf {\bibinfo
  {volume} {103}},\ \bibinfo {pages} {247801} (\bibinfo {year}
  {2009})}\BibitemShut {NoStop}%
\bibitem [{\citenamefont {H{\'e}not}\ \emph {et~al.}(2018)\citenamefont
  {H{\'e}not}, \citenamefont {Grzelka}, \citenamefont {Zhang}, \citenamefont
  {Mariot}, \citenamefont {Antoniuk}, \citenamefont {Drockenmuller},
  \citenamefont {L{\'e}ger},\ and\ \citenamefont {Restagno}}]{Henot2018}%
  \BibitemOpen
  \bibfield  {author} {\bibinfo {author} {\bibfnamefont {M.}~\bibnamefont
  {H{\'e}not}}, \bibinfo {author} {\bibfnamefont {M.}~\bibnamefont {Grzelka}},
  \bibinfo {author} {\bibfnamefont {J.}~\bibnamefont {Zhang}}, \bibinfo
  {author} {\bibfnamefont {S.}~\bibnamefont {Mariot}}, \bibinfo {author}
  {\bibfnamefont {I.}~\bibnamefont {Antoniuk}}, \bibinfo {author}
  {\bibfnamefont {E.}~\bibnamefont {Drockenmuller}}, \bibinfo {author}
  {\bibfnamefont {L.}~\bibnamefont {L{\'e}ger}},\ and\ \bibinfo {author}
  {\bibfnamefont {F.}~\bibnamefont {Restagno}},\ }\bibfield  {title} {\bibinfo
  {title} {Temperature-{{Controlled Slip}} of {{Polymer Melts}} on {{Ideal
  Substrates}}},\ }\href {https://doi.org/10.1103/PhysRevLett.121.177802}
  {\bibfield  {journal} {\bibinfo  {journal} {Physical Review Letters}\
  }\textbf {\bibinfo {volume} {121}},\ \bibinfo {pages} {177802} (\bibinfo
  {year} {2018})}\BibitemShut {NoStop}%
\bibitem [{\citenamefont {Ilton}\ \emph {et~al.}(2018)\citenamefont {Ilton},
  \citenamefont {Salez}, \citenamefont {Fowler}, \citenamefont {Rivetti},
  \citenamefont {Aly}, \citenamefont {Benzaquen}, \citenamefont {McGraw},
  \citenamefont {Rapha{\"e}l}, \citenamefont {{Dalnoki-Veress}},\ and\
  \citenamefont {B{\"a}umchen}}]{Ilton2018}%
  \BibitemOpen
  \bibfield  {author} {\bibinfo {author} {\bibfnamefont {M.}~\bibnamefont
  {Ilton}}, \bibinfo {author} {\bibfnamefont {T.}~\bibnamefont {Salez}},
  \bibinfo {author} {\bibfnamefont {P.~D.}\ \bibnamefont {Fowler}}, \bibinfo
  {author} {\bibfnamefont {M.}~\bibnamefont {Rivetti}}, \bibinfo {author}
  {\bibfnamefont {M.}~\bibnamefont {Aly}}, \bibinfo {author} {\bibfnamefont
  {M.}~\bibnamefont {Benzaquen}}, \bibinfo {author} {\bibfnamefont {J.~D.}\
  \bibnamefont {McGraw}}, \bibinfo {author} {\bibfnamefont {E.}~\bibnamefont
  {Rapha{\"e}l}}, \bibinfo {author} {\bibfnamefont {K.}~\bibnamefont
  {{Dalnoki-Veress}}},\ and\ \bibinfo {author} {\bibfnamefont {O.}~\bibnamefont
  {B{\"a}umchen}},\ }\bibfield  {title} {\bibinfo {title} {Adsorption-induced
  slip inhibition for polymer melts on ideal substrates},\ }\href
  {https://doi.org/10.1038/s41467-018-03610-4} {\bibfield  {journal} {\bibinfo
  {journal} {Nature Communications}\ }\textbf {\bibinfo {volume} {9}},\
  \bibinfo {pages} {1172} (\bibinfo {year} {2018})}\BibitemShut {NoStop}%
\bibitem [{\citenamefont {Barnes}(1995)}]{BarnesReview1995}%
  \BibitemOpen
  \bibfield  {author} {\bibinfo {author} {\bibfnamefont {H.~A.}\ \bibnamefont
  {Barnes}},\ }\bibfield  {title} {\bibinfo {title} {A review of the slip (wall
  depletion) of polymer solutions, emulsions and particle suspensions in
  viscometers: Its cause, character, and cure},\ }\href
  {https://doi.org/10.1016/0377-0257(94)01282-M} {\bibfield  {journal}
  {\bibinfo  {journal} {Journal of Non-Newtonian Fluid Mechanics}\ }\textbf
  {\bibinfo {volume} {56}},\ \bibinfo {pages} {221} (\bibinfo {year}
  {1995})}\BibitemShut {NoStop}%
\bibitem [{\citenamefont {Cuenca}\ and\ \citenamefont
  {Bodiguel}(2013)}]{Cuenca2013}%
  \BibitemOpen
  \bibfield  {author} {\bibinfo {author} {\bibfnamefont {A.}~\bibnamefont
  {Cuenca}}\ and\ \bibinfo {author} {\bibfnamefont {H.}~\bibnamefont
  {Bodiguel}},\ }\bibfield  {title} {\bibinfo {title} {Submicron {{Flow}} of
  {{Polymer Solutions}}: {{Slippage Reduction}} due to {{Confinement}}},\
  }\href {https://doi.org/10.1103/PhysRevLett.110.108304} {\bibfield  {journal}
  {\bibinfo  {journal} {Physical Review Letters}\ }\textbf {\bibinfo {volume}
  {110}},\ \bibinfo {pages} {108304} (\bibinfo {year} {2013})}\BibitemShut
  {NoStop}%
\bibitem [{\citenamefont {Barraud}\ \emph {et~al.}(2019)\citenamefont
  {Barraud}, \citenamefont {Cross}, \citenamefont {Picard}, \citenamefont
  {Restagno}, \citenamefont {L{\'e}ger},\ and\ \citenamefont
  {Charlaix}}]{Barraud2019}%
  \BibitemOpen
  \bibfield  {author} {\bibinfo {author} {\bibfnamefont {C.}~\bibnamefont
  {Barraud}}, \bibinfo {author} {\bibfnamefont {B.}~\bibnamefont {Cross}},
  \bibinfo {author} {\bibfnamefont {C.}~\bibnamefont {Picard}}, \bibinfo
  {author} {\bibfnamefont {F.}~\bibnamefont {Restagno}}, \bibinfo {author}
  {\bibfnamefont {L.}~\bibnamefont {L{\'e}ger}},\ and\ \bibinfo {author}
  {\bibfnamefont {E.}~\bibnamefont {Charlaix}},\ }\bibfield  {title} {\bibinfo
  {title} {Large slippage and depletion layer at the polyelectrolyte/solid
  interface},\ }\href {https://doi.org/10.1039/C9SM00910H} {\bibfield
  {journal} {\bibinfo  {journal} {Soft Matter}\ }\textbf {\bibinfo {volume}
  {15}},\ \bibinfo {pages} {6308} (\bibinfo {year} {2019})}\BibitemShut
  {NoStop}%
\bibitem [{\citenamefont {Cross}\ \emph {et~al.}(2018)\citenamefont {Cross},
  \citenamefont {Barraud}, \citenamefont {Picard}, \citenamefont {L{\'e}ger},
  \citenamefont {Restagno},\ and\ \citenamefont {Charlaix}}]{Cross2018}%
  \BibitemOpen
  \bibfield  {author} {\bibinfo {author} {\bibfnamefont {B.}~\bibnamefont
  {Cross}}, \bibinfo {author} {\bibfnamefont {C.}~\bibnamefont {Barraud}},
  \bibinfo {author} {\bibfnamefont {C.}~\bibnamefont {Picard}}, \bibinfo
  {author} {\bibfnamefont {L.}~\bibnamefont {L{\'e}ger}}, \bibinfo {author}
  {\bibfnamefont {F.}~\bibnamefont {Restagno}},\ and\ \bibinfo {author}
  {\bibfnamefont {{\'E}.}~\bibnamefont {Charlaix}},\ }\bibfield  {title}
  {\bibinfo {title} {Wall slip of complex fluids: {{Interfacial}} friction
  versus slip length},\ }\href {https://doi.org/10.1103/PhysRevFluids.3.062001}
  {\bibfield  {journal} {\bibinfo  {journal} {Physical Review Fluids}\ }\textbf
  {\bibinfo {volume} {3}},\ \bibinfo {pages} {062001} (\bibinfo {year}
  {2018})}\BibitemShut {NoStop}%
\bibitem [{\citenamefont {Park}\ \emph {et~al.}(2019)\citenamefont {Park},
  \citenamefont {Shakya},\ and\ \citenamefont {King}}]{Park2019}%
  \BibitemOpen
  \bibfield  {author} {\bibinfo {author} {\bibfnamefont {S.~J.}\ \bibnamefont
  {Park}}, \bibinfo {author} {\bibfnamefont {A.}~\bibnamefont {Shakya}},\ and\
  \bibinfo {author} {\bibfnamefont {J.~T.}\ \bibnamefont {King}},\ }\bibfield
  {title} {\bibinfo {title} {Depletion layer dynamics of polyelectrolyte
  solutions under {Poiseuille} flow},\ }\href
  {https://doi.org/10.1073/pnas.1900623116} {\bibfield  {journal} {\bibinfo
  {journal} {Proceedings of the National Academy of Sciences}\ }\textbf
  {\bibinfo {volume} {116}},\ \bibinfo {pages} {16256} (\bibinfo {year}
  {2019})}\BibitemShut {NoStop}%
\bibitem [{\citenamefont {Grzelka}\ \emph {et~al.}(2021)\citenamefont
  {Grzelka}, \citenamefont {Antoniuk}, \citenamefont {Drockenmuller},
  \citenamefont {Chennevi{\`e}re}, \citenamefont {L{\'e}ger},\ and\
  \citenamefont {Restagno}}]{Grzelka2021}%
  \BibitemOpen
  \bibfield  {author} {\bibinfo {author} {\bibfnamefont {M.}~\bibnamefont
  {Grzelka}}, \bibinfo {author} {\bibfnamefont {I.}~\bibnamefont {Antoniuk}},
  \bibinfo {author} {\bibfnamefont {{\'E}.}~\bibnamefont {Drockenmuller}},
  \bibinfo {author} {\bibfnamefont {A.}~\bibnamefont {Chennevi{\`e}re}},
  \bibinfo {author} {\bibfnamefont {L.}~\bibnamefont {L{\'e}ger}},\ and\
  \bibinfo {author} {\bibfnamefont {F.}~\bibnamefont {Restagno}},\ }\bibfield
  {title} {\bibinfo {title} {Slip and {Friction} {Mechanisms} at {Polymer}
  {Semi}-{Dilute} {Solutions}/{Solid} {Interfaces}},\ }\href
  {https://doi.org/10.1021/acs.macromol.0c02804} {\bibfield  {journal}
  {\bibinfo  {journal} {Macromolecules}\ }\textbf {\bibinfo {volume} {54}},\
  \bibinfo {pages} {4910} (\bibinfo {year} {2021})}\BibitemShut {NoStop}%
\bibitem [{\citenamefont {Guyard}\ \emph {et~al.}(2021)\citenamefont {Guyard},
  \citenamefont {Vilquin}, \citenamefont {Sanson}, \citenamefont {Jouenne},
  \citenamefont {Restagno},\ and\ \citenamefont {McGraw}}]{Guyard2021}%
  \BibitemOpen
  \bibfield  {author} {\bibinfo {author} {\bibfnamefont {G.}~\bibnamefont
  {Guyard}}, \bibinfo {author} {\bibfnamefont {A.}~\bibnamefont {Vilquin}},
  \bibinfo {author} {\bibfnamefont {N.}~\bibnamefont {Sanson}}, \bibinfo
  {author} {\bibfnamefont {S.}~\bibnamefont {Jouenne}}, \bibinfo {author}
  {\bibfnamefont {F.}~\bibnamefont {Restagno}},\ and\ \bibinfo {author}
  {\bibfnamefont {J.~D.}\ \bibnamefont {McGraw}},\ }\bibfield  {title}
  {{\selectlanguage {english}\bibinfo {title} {Near-surface rheology and
  hydrodynamic boundary condition of semi-dilute polymer solutions}},\ }\href
  {https://doi.org/10.1039/D0SM02116D} {\bibfield  {journal} {\bibinfo
  {journal} {Soft Matter}\ }\textbf {\bibinfo {volume} {17}},\ \bibinfo {pages}
  {3765} (\bibinfo {year} {2021})}\BibitemShut {NoStop}%
\bibitem [{\citenamefont {Salmon}\ \emph {et~al.}(2003)\citenamefont {Salmon},
  \citenamefont {B{\'e}cu}, \citenamefont {Manneville},\ and\ \citenamefont
  {Colin}}]{salmon_towards_2003}%
  \BibitemOpen
  \bibfield  {author} {\bibinfo {author} {\bibfnamefont {J.-B.}\ \bibnamefont
  {Salmon}}, \bibinfo {author} {\bibfnamefont {L.}~\bibnamefont {B{\'e}cu}},
  \bibinfo {author} {\bibfnamefont {S.}~\bibnamefont {Manneville}},\ and\
  \bibinfo {author} {\bibfnamefont {A.}~\bibnamefont {Colin}},\ }\bibfield
  {title} {\bibinfo {title} {Towards local rheology of emulsions under
  {Couette} flow using {Dynamic} {Light} {Scattering}},\ }\href@noop {}
  {\bibfield  {journal} {\bibinfo  {journal} {Eur. Phys. J. E}\ }\textbf
  {\bibinfo {volume} {10}},\ \bibinfo {pages} {209} (\bibinfo {year}
  {2003})}\BibitemShut {NoStop}%
\bibitem [{\citenamefont {Goyon}\ \emph {et~al.}(2008)\citenamefont {Goyon},
  \citenamefont {Colin}, \citenamefont {Ovarlez}, \citenamefont {Ajdari},\ and\
  \citenamefont {Bocquet}}]{Goyon2008}%
  \BibitemOpen
  \bibfield  {author} {\bibinfo {author} {\bibfnamefont {J.}~\bibnamefont
  {Goyon}}, \bibinfo {author} {\bibfnamefont {A.}~\bibnamefont {Colin}},
  \bibinfo {author} {\bibfnamefont {G.}~\bibnamefont {Ovarlez}}, \bibinfo
  {author} {\bibfnamefont {A.}~\bibnamefont {Ajdari}},\ and\ \bibinfo {author}
  {\bibfnamefont {L.}~\bibnamefont {Bocquet}},\ }\bibfield  {title} {\bibinfo
  {title} {Spatial cooperativity in soft glassy flows},\ }\href
  {https://doi.org/10.1038/nature07026} {\bibfield  {journal} {\bibinfo
  {journal} {Nature}\ }\textbf {\bibinfo {volume} {454}},\ \bibinfo {pages}
  {84} (\bibinfo {year} {2008})}\BibitemShut {NoStop}%
\bibitem [{\citenamefont {Huerre}\ \emph {et~al.}(2015)\citenamefont {Huerre},
  \citenamefont {Theodoly}, \citenamefont {Leshansky}, \citenamefont
  {Valignat}, \citenamefont {Cantat},\ and\ \citenamefont
  {Jullien}}]{Huerre2015}%
  \BibitemOpen
  \bibfield  {author} {\bibinfo {author} {\bibfnamefont {A.}~\bibnamefont
  {Huerre}}, \bibinfo {author} {\bibfnamefont {O.}~\bibnamefont {Theodoly}},
  \bibinfo {author} {\bibfnamefont {A.~M.}\ \bibnamefont {Leshansky}}, \bibinfo
  {author} {\bibfnamefont {M.-P.}\ \bibnamefont {Valignat}}, \bibinfo {author}
  {\bibfnamefont {I.}~\bibnamefont {Cantat}},\ and\ \bibinfo {author}
  {\bibfnamefont {M.-C.}\ \bibnamefont {Jullien}},\ }\bibfield  {title}
  {{\selectlanguage {english}\bibinfo {title} {Droplets in {Microchannels}:
  {Dynamical} {Properties} of the {Lubrication} {Film}}},\ }\href
  {https://doi.org/10.1103/PhysRevLett.115.064501} {\bibfield  {journal}
  {\bibinfo  {journal} {Physical Review Letters}\ }\textbf {\bibinfo {volume}
  {115}},\ \bibinfo {pages} {064501} (\bibinfo {year} {2015})}\BibitemShut
  {NoStop}%
\bibitem [{\citenamefont {Mansard}\ \emph
  {et~al.}(2014{\natexlab{a}})\citenamefont {Mansard}, \citenamefont
  {Bocquet},\ and\ \citenamefont {Colin}}]{Mansard2014}%
  \BibitemOpen
  \bibfield  {author} {\bibinfo {author} {\bibfnamefont {V.}~\bibnamefont
  {Mansard}}, \bibinfo {author} {\bibfnamefont {L.}~\bibnamefont {Bocquet}},\
  and\ \bibinfo {author} {\bibfnamefont {A.}~\bibnamefont {Colin}},\ }\bibfield
   {title} {\bibinfo {title} {Boundary conditions for soft glassy flows:
  Slippage and surface fluidization},\ }\href
  {https://doi.org/10.1039/C4SM00230J} {\bibfield  {journal} {\bibinfo
  {journal} {Soft Matter}\ }\textbf {\bibinfo {volume} {10}},\ \bibinfo {pages}
  {6984} (\bibinfo {year} {2014}{\natexlab{a}})}\BibitemShut {NoStop}%
\bibitem [{\citenamefont {Denkov}\ \emph {et~al.}(2005)\citenamefont {Denkov},
  \citenamefont {Subramanian}, \citenamefont {Gurovich},\ and\ \citenamefont
  {Lips}}]{denkov_wall_2005}%
  \BibitemOpen
  \bibfield  {author} {\bibinfo {author} {\bibfnamefont {N.~D.}\ \bibnamefont
  {Denkov}}, \bibinfo {author} {\bibfnamefont {V.}~\bibnamefont {Subramanian}},
  \bibinfo {author} {\bibfnamefont {D.}~\bibnamefont {Gurovich}},\ and\
  \bibinfo {author} {\bibfnamefont {A.}~\bibnamefont {Lips}},\ }\bibfield
  {title} {\bibinfo {title} {Wall slip and viscous dissipation in sheared
  foams: {Effect} of surface mobility},\ }\href@noop {} {\bibfield  {journal}
  {\bibinfo  {journal} {Colloids and Surfaces A}\ }\textbf {\bibinfo {volume}
  {263}},\ \bibinfo {pages} {129} (\bibinfo {year} {2005})}\BibitemShut
  {NoStop}%
\bibitem [{\citenamefont {Marze}\ \emph {et~al.}(2008)\citenamefont {Marze},
  \citenamefont {Langevin},\ and\ \citenamefont
  {Saint-Jalmes}}]{marze_aqueous_2008}%
  \BibitemOpen
  \bibfield  {author} {\bibinfo {author} {\bibfnamefont {S.}~\bibnamefont
  {Marze}}, \bibinfo {author} {\bibfnamefont {D.}~\bibnamefont {Langevin}},\
  and\ \bibinfo {author} {\bibfnamefont {A.}~\bibnamefont {Saint-Jalmes}},\
  }\bibfield  {title} {\bibinfo {title} {Aqueous foam slip and shear regimes
  determined by rheometry and multiple light scattering},\ }\href
  {https://doi.org/10.1122/1.2952510} {\bibfield  {journal} {\bibinfo
  {journal} {Journal of Rheology}\ }\textbf {\bibinfo {volume} {52}},\ \bibinfo
  {pages} {1091} (\bibinfo {year} {2008})}\BibitemShut {NoStop}%
\bibitem [{\citenamefont {Cantat}(2013)}]{cantat_liquid_2013}%
  \BibitemOpen
  \bibfield  {author} {\bibinfo {author} {\bibfnamefont {I.}~\bibnamefont
  {Cantat}},\ }\bibfield  {title} {\bibinfo {title} {Liquid meniscus friction
  on a wet plate: {Bubbles}, lamellae and foams},\ }\href@noop {} {\bibfield
  {journal} {\bibinfo  {journal} {Phys. Fluids}\ }\textbf {\bibinfo {volume}
  {25}},\ \bibinfo {pages} {031303} (\bibinfo {year} {2013})}\BibitemShut
  {NoStop}%
\bibitem [{\citenamefont {Le~Merrer}\ \emph {et~al.}(2015)\citenamefont
  {Le~Merrer}, \citenamefont {Lespiat}, \citenamefont {H{\"o}hler},\ and\
  \citenamefont {Cohen-Addad}}]{le_merrer_linear_2015}%
  \BibitemOpen
  \bibfield  {author} {\bibinfo {author} {\bibfnamefont {M.}~\bibnamefont
  {Le~Merrer}}, \bibinfo {author} {\bibfnamefont {R.}~\bibnamefont {Lespiat}},
  \bibinfo {author} {\bibfnamefont {R.}~\bibnamefont {H{\"o}hler}},\ and\
  \bibinfo {author} {\bibfnamefont {S.}~\bibnamefont {Cohen-Addad}},\
  }\bibfield  {title} {\bibinfo {title} {Linear and non-linear wall friction of
  wet foams},\ }\href {https://doi.org/10.1039/C4SM01557F} {\bibfield
  {journal} {\bibinfo  {journal} {Soft Matter}\ }\textbf {\bibinfo {volume}
  {11}},\ \bibinfo {pages} {368} (\bibinfo {year} {2015})}\BibitemShut
  {NoStop}%
\bibitem [{\citenamefont {Meeker}\ \emph
  {et~al.}(2004{\natexlab{a}})\citenamefont {Meeker}, \citenamefont
  {Bonnecaze},\ and\ \citenamefont {Cloitre}}]{Meeker2004}%
  \BibitemOpen
  \bibfield  {author} {\bibinfo {author} {\bibfnamefont {S.~P.}\ \bibnamefont
  {Meeker}}, \bibinfo {author} {\bibfnamefont {R.~T.}\ \bibnamefont
  {Bonnecaze}},\ and\ \bibinfo {author} {\bibfnamefont {M.}~\bibnamefont
  {Cloitre}},\ }\bibfield  {title} {{\selectlanguage {english}\bibinfo {title}
  {Slip and flow in pastes of soft particles: {Direct} observation and
  rheology}},\ }\href {https://doi.org/10.1122/1.1795171} {\bibfield  {journal}
  {\bibinfo  {journal} {Journal of Rheology}\ }\textbf {\bibinfo {volume}
  {48}},\ \bibinfo {pages} {1295} (\bibinfo {year}
  {2004}{\natexlab{a}})}\BibitemShut {NoStop}%
\bibitem [{\citenamefont {Meeker}\ \emph
  {et~al.}(2004{\natexlab{b}})\citenamefont {Meeker}, \citenamefont
  {Bonnecaze},\ and\ \citenamefont {Cloitre}}]{Meeker2004PRL}%
  \BibitemOpen
  \bibfield  {author} {\bibinfo {author} {\bibfnamefont {S.~P.}\ \bibnamefont
  {Meeker}}, \bibinfo {author} {\bibfnamefont {R.~T.}\ \bibnamefont
  {Bonnecaze}},\ and\ \bibinfo {author} {\bibfnamefont {M.}~\bibnamefont
  {Cloitre}},\ }\bibfield  {title} {\bibinfo {title} {Slip and {Flow} in {Soft}
  {Particle} {Pastes}},\ }\href {https://doi.org/10.1103/PhysRevLett.92.198302}
  {\bibfield  {journal} {\bibinfo  {journal} {Physical Review Letters}\
  }\textbf {\bibinfo {volume} {92}},\ \bibinfo {pages} {198302} (\bibinfo
  {year} {2004}{\natexlab{b}})}\BibitemShut {NoStop}%
\bibitem [{\citenamefont {Divoux}\ \emph {et~al.}(2015)\citenamefont {Divoux},
  \citenamefont {Lapeyre}, \citenamefont {Ravaine},\ and\ \citenamefont
  {Manneville}}]{Divoux2015}%
  \BibitemOpen
  \bibfield  {author} {\bibinfo {author} {\bibfnamefont {T.}~\bibnamefont
  {Divoux}}, \bibinfo {author} {\bibfnamefont {V.}~\bibnamefont {Lapeyre}},
  \bibinfo {author} {\bibfnamefont {V.}~\bibnamefont {Ravaine}},\ and\ \bibinfo
  {author} {\bibfnamefont {S.}~\bibnamefont {Manneville}},\ }\bibfield  {title}
  {{\selectlanguage {english}\bibinfo {title} {Wall slip across the jamming
  transition of soft thermoresponsive particles}},\ }\href
  {https://doi.org/10.1103/PhysRevE.92.060301} {\bibfield  {journal} {\bibinfo
  {journal} {Physical Review E}\ }\textbf {\bibinfo {volume} {92}},\ \bibinfo
  {pages} {060301} (\bibinfo {year} {2015})}\BibitemShut {NoStop}%
\bibitem [{\citenamefont {Jalaal}\ \emph {et~al.}(2015)\citenamefont {Jalaal},
  \citenamefont {Balmforth},\ and\ \citenamefont {Stoeber}}]{Jalaal2015}%
  \BibitemOpen
  \bibfield  {author} {\bibinfo {author} {\bibfnamefont {M.}~\bibnamefont
  {Jalaal}}, \bibinfo {author} {\bibfnamefont {N.~J.}\ \bibnamefont
  {Balmforth}},\ and\ \bibinfo {author} {\bibfnamefont {B.}~\bibnamefont
  {Stoeber}},\ }\bibfield  {title} {\bibinfo {title} {Slip of {{Spreading
  Viscoplastic Droplets}}},\ }\href
  {https://doi.org/10.1021/acs.langmuir.5b02353} {\bibfield  {journal}
  {\bibinfo  {journal} {Langmuir}\ }\textbf {\bibinfo {volume} {31}},\ \bibinfo
  {pages} {12071} (\bibinfo {year} {2015})}\BibitemShut {NoStop}%
\bibitem [{\citenamefont {Zhang}\ \emph {et~al.}(2017)\citenamefont {Zhang},
  \citenamefont {Lorenceau}, \citenamefont {Basset}, \citenamefont {Bourouina},
  \citenamefont {Rouyer}, \citenamefont {Goyon},\ and\ \citenamefont
  {Coussot}}]{Zhang2017}%
  \BibitemOpen
  \bibfield  {author} {\bibinfo {author} {\bibfnamefont {X.}~\bibnamefont
  {Zhang}}, \bibinfo {author} {\bibfnamefont {E.}~\bibnamefont {Lorenceau}},
  \bibinfo {author} {\bibfnamefont {P.}~\bibnamefont {Basset}}, \bibinfo
  {author} {\bibfnamefont {T.}~\bibnamefont {Bourouina}}, \bibinfo {author}
  {\bibfnamefont {F.}~\bibnamefont {Rouyer}}, \bibinfo {author} {\bibfnamefont
  {J.}~\bibnamefont {Goyon}},\ and\ \bibinfo {author} {\bibfnamefont
  {P.}~\bibnamefont {Coussot}},\ }\bibfield  {title} {{\selectlanguage
  {english}\bibinfo {title} {Wall {Slip} of {Soft}-{Jammed} {Systems}: {A}
  {Generic} {Simple} {Shear} {Process}}},\ }\href
  {https://doi.org/10.1103/PhysRevLett.119.208004} {\bibfield  {journal}
  {\bibinfo  {journal} {Physical Review Letters}\ }\textbf {\bibinfo {volume}
  {119}},\ \bibinfo {pages} {208004} (\bibinfo {year} {2017})}\BibitemShut
  {NoStop}%
\bibitem [{\citenamefont {P{\'e}m{\'e}ja}\ \emph {et~al.}(2019)\citenamefont
  {P{\'e}m{\'e}ja}, \citenamefont {G{\'e}raud}, \citenamefont {Barentin},\ and\
  \citenamefont {Le~Merrer}}]{Pemeja2019}%
  \BibitemOpen
  \bibfield  {author} {\bibinfo {author} {\bibfnamefont {J.}~\bibnamefont
  {P{\'e}m{\'e}ja}}, \bibinfo {author} {\bibfnamefont {B.}~\bibnamefont
  {G{\'e}raud}}, \bibinfo {author} {\bibfnamefont {C.}~\bibnamefont
  {Barentin}},\ and\ \bibinfo {author} {\bibfnamefont {M.}~\bibnamefont
  {Le~Merrer}},\ }\bibfield  {title} {{\selectlanguage {english}\bibinfo
  {title} {Wall slip regimes in jammed suspensions of soft microgels}},\ }\href
  {https://doi.org/10.1103/PhysRevFluids.4.033301} {\bibfield  {journal}
  {\bibinfo  {journal} {Physical Review Fluids}\ }\textbf {\bibinfo {volume}
  {4}},\ \bibinfo {pages} {033301} (\bibinfo {year} {2019})}\BibitemShut
  {NoStop}%
\bibitem [{\citenamefont {Coussot}(2007)}]{Coussot2007}%
  \BibitemOpen
  \bibfield  {author} {\bibinfo {author} {\bibfnamefont {P.}~\bibnamefont
  {Coussot}},\ }\bibfield  {title} {\bibinfo {title} {Rheophysics of pastes: A
  review of microscopic modelling approaches},\ }\href
  {https://doi.org/10.1039/B611021P} {\bibfield  {journal} {\bibinfo  {journal}
  {Soft Matter}\ }\textbf {\bibinfo {volume} {3}},\ \bibinfo {pages} {528}
  (\bibinfo {year} {2007})}\BibitemShut {NoStop}%
\bibitem [{\citenamefont {Bonn}\ \emph {et~al.}(2017)\citenamefont {Bonn},
  \citenamefont {Denn}, \citenamefont {Berthier}, \citenamefont {Divoux},\ and\
  \citenamefont {Manneville}}]{Bonn2017}%
  \BibitemOpen
  \bibfield  {author} {\bibinfo {author} {\bibfnamefont {D.}~\bibnamefont
  {Bonn}}, \bibinfo {author} {\bibfnamefont {M.~M.}\ \bibnamefont {Denn}},
  \bibinfo {author} {\bibfnamefont {L.}~\bibnamefont {Berthier}}, \bibinfo
  {author} {\bibfnamefont {T.}~\bibnamefont {Divoux}},\ and\ \bibinfo {author}
  {\bibfnamefont {S.}~\bibnamefont {Manneville}},\ }\bibfield  {title}
  {\bibinfo {title} {Yield stress materials in soft condensed matter},\ }\href
  {https://doi.org/10.1103/RevModPhys.89.035005} {\bibfield  {journal}
  {\bibinfo  {journal} {Reviews of Modern Physics}\ }\textbf {\bibinfo {volume}
  {89}},\ \bibinfo {pages} {035005} (\bibinfo {year} {2017})}\BibitemShut
  {NoStop}%
\bibitem [{\citenamefont {Cloitre}\ and\ \citenamefont
  {Bonnecaze}(2017)}]{Cloitre2017}%
  \BibitemOpen
  \bibfield  {author} {\bibinfo {author} {\bibfnamefont {M.}~\bibnamefont
  {Cloitre}}\ and\ \bibinfo {author} {\bibfnamefont {R.~T.}\ \bibnamefont
  {Bonnecaze}},\ }\bibfield  {title} {{\selectlanguage {english}\bibinfo
  {title} {A review on wall slip in high solid dispersions}},\ }\href
  {https://doi.org/10.1007/s00397-017-1002-7} {\bibfield  {journal} {\bibinfo
  {journal} {Rheologica Acta}\ }\textbf {\bibinfo {volume} {56}},\ \bibinfo
  {pages} {283} (\bibinfo {year} {2017})}\BibitemShut {NoStop}%
\bibitem [{\citenamefont {Skotheim}\ and\ \citenamefont
  {Mahadevan}(2004)}]{Skotheim2004}%
  \BibitemOpen
  \bibfield  {author} {\bibinfo {author} {\bibfnamefont {J.~M.}\ \bibnamefont
  {Skotheim}}\ and\ \bibinfo {author} {\bibfnamefont {L.}~\bibnamefont
  {Mahadevan}},\ }\bibfield  {title} {\bibinfo {title} {Soft {{Lubrication}}},\
  }\href {https://doi.org/10.1103/PhysRevLett.92.245509} {\bibfield  {journal}
  {\bibinfo  {journal} {Physical Review Letters}\ }\textbf {\bibinfo {volume}
  {92}},\ \bibinfo {pages} {245509} (\bibinfo {year} {2004})}\BibitemShut
  {NoStop}%
\bibitem [{\citenamefont {Rubinstein}\ and\ \citenamefont
  {Colby}(2003)}]{Rubinstein2003}%
  \BibitemOpen
  \bibfield  {author} {\bibinfo {author} {\bibfnamefont {M.}~\bibnamefont
  {Rubinstein}}\ and\ \bibinfo {author} {\bibfnamefont {R.~H.}\ \bibnamefont
  {Colby}},\ }\href@noop {} {\emph {\bibinfo {title} {Polymer {{Physics}}}}}\
  (\bibinfo  {publisher} {Oxford University Press},\ \bibinfo {address}
  {Oxford; New York},\ \bibinfo {year} {2003})\BibitemShut {NoStop}%
\bibitem [{\citenamefont {Joanny}\ \emph {et~al.}(1979)\citenamefont {Joanny},
  \citenamefont {Leibler},\ and\ \citenamefont {De~Gennes}}]{Joanny1979}%
  \BibitemOpen
  \bibfield  {author} {\bibinfo {author} {\bibfnamefont {J.~F.}\ \bibnamefont
  {Joanny}}, \bibinfo {author} {\bibfnamefont {L.}~\bibnamefont {Leibler}},\
  and\ \bibinfo {author} {\bibfnamefont {P.~G.}\ \bibnamefont {De~Gennes}},\
  }\bibfield  {title} {\bibinfo {title} {Effects of polymer solutions on
  colloid stability},\ }\href {https://doi.org/10.1002/pol.1979.180170615}
  {\bibfield  {journal} {\bibinfo  {journal} {Journal of Polymer Science:
  Polymer Physics Edition}\ }\textbf {\bibinfo {volume} {17}},\ \bibinfo
  {pages} {1073} (\bibinfo {year} {1979})}\BibitemShut {NoStop}%
\bibitem [{\citenamefont {Dobrynin}\ \emph {et~al.}(1995)\citenamefont
  {Dobrynin}, \citenamefont {Colby},\ and\ \citenamefont
  {Rubinstein}}]{Dobrynin1995}%
  \BibitemOpen
  \bibfield  {author} {\bibinfo {author} {\bibfnamefont {A.~V.}\ \bibnamefont
  {Dobrynin}}, \bibinfo {author} {\bibfnamefont {R.~H.}\ \bibnamefont
  {Colby}},\ and\ \bibinfo {author} {\bibfnamefont {M.}~\bibnamefont
  {Rubinstein}},\ }\bibfield  {title} {\bibinfo {title} {Scaling {{Theory}} of
  {{Polyelectrolyte Solutions}}},\ }\href {https://doi.org/10.1021/ma00110a021}
  {\bibfield  {journal} {\bibinfo  {journal} {Macromolecules}\ }\textbf
  {\bibinfo {volume} {28}},\ \bibinfo {pages} {1859} (\bibinfo {year}
  {1995})}\BibitemShut {NoStop}%
\bibitem [{\citenamefont {Lafon}\ \emph {et~al.}(2025)\citenamefont {Lafon},
  \citenamefont {Outerelo~Corvo}, \citenamefont {Grzelka}, \citenamefont
  {H{\'e}lary}, \citenamefont {Gutfreund}, \citenamefont {L{\'e}ger},
  \citenamefont {Chennevi{\`e}re},\ and\ \citenamefont {Restagno}}]{Lafon2025}%
  \BibitemOpen
  \bibfield  {author} {\bibinfo {author} {\bibfnamefont {S.}~\bibnamefont
  {Lafon}}, \bibinfo {author} {\bibfnamefont {T.}~\bibnamefont
  {Outerelo~Corvo}}, \bibinfo {author} {\bibfnamefont {M.}~\bibnamefont
  {Grzelka}}, \bibinfo {author} {\bibfnamefont {A.}~\bibnamefont {H{\'e}lary}},
  \bibinfo {author} {\bibfnamefont {P.}~\bibnamefont {Gutfreund}}, \bibinfo
  {author} {\bibfnamefont {L.}~\bibnamefont {L{\'e}ger}}, \bibinfo {author}
  {\bibfnamefont {A.}~\bibnamefont {Chennevi{\`e}re}},\ and\ \bibinfo {author}
  {\bibfnamefont {F.}~\bibnamefont {Restagno}},\ }\bibfield  {title} {\bibinfo
  {title} {Near-{{Surface Concentration Profile}} of {{Sheared Semidilute
  Polymer Solutions}}},\ }\href {https://doi.org/10.1021/acs.langmuir.4c04005}
  {\bibfield  {journal} {\bibinfo  {journal} {Langmuir}\ }\textbf {\bibinfo
  {volume} {41}},\ \bibinfo {pages} {1716} (\bibinfo {year}
  {2025})}\BibitemShut {NoStop}%
\bibitem [{\citenamefont {Denkov}\ \emph {et~al.}(2006)\citenamefont {Denkov},
  \citenamefont {Tcholakova}, \citenamefont {Golemanov}, \citenamefont
  {Subramanian},\ and\ \citenamefont {Lips}}]{denkov_foam_2006}%
  \BibitemOpen
  \bibfield  {author} {\bibinfo {author} {\bibfnamefont {N.~D.}\ \bibnamefont
  {Denkov}}, \bibinfo {author} {\bibfnamefont {S.}~\bibnamefont {Tcholakova}},
  \bibinfo {author} {\bibfnamefont {K.}~\bibnamefont {Golemanov}}, \bibinfo
  {author} {\bibfnamefont {V.}~\bibnamefont {Subramanian}},\ and\ \bibinfo
  {author} {\bibfnamefont {A.}~\bibnamefont {Lips}},\ }\bibfield  {title}
  {\bibinfo {title} {Foam wall friction: {Effect} of air volume fraction for
  tangentially immobile bubble surface},\ }\href@noop {} {\bibfield  {journal}
  {\bibinfo  {journal} {Colloids and Surfaces A}\ }\textbf {\bibinfo {volume}
  {282}},\ \bibinfo {pages} {329} (\bibinfo {year} {2006})}\BibitemShut
  {NoStop}%
\bibitem [{\citenamefont {Reichert}\ \emph {et~al.}(2018)\citenamefont
  {Reichert}, \citenamefont {Huerre}, \citenamefont {Theodoly}, \citenamefont
  {Valignat}, \citenamefont {Cantat},\ and\ \citenamefont
  {Jullien}}]{Reichert2018}%
  \BibitemOpen
  \bibfield  {author} {\bibinfo {author} {\bibfnamefont {B.}~\bibnamefont
  {Reichert}}, \bibinfo {author} {\bibfnamefont {A.}~\bibnamefont {Huerre}},
  \bibinfo {author} {\bibfnamefont {O.}~\bibnamefont {Theodoly}}, \bibinfo
  {author} {\bibfnamefont {M.-P.}\ \bibnamefont {Valignat}}, \bibinfo {author}
  {\bibfnamefont {I.}~\bibnamefont {Cantat}},\ and\ \bibinfo {author}
  {\bibfnamefont {M.-C.}\ \bibnamefont {Jullien}},\ }\bibfield  {title}
  {\bibinfo {title} {Topography of the lubrication film under a pancake droplet
  travelling in a {{Hele-Shaw}} cell},\ }\href
  {https://doi.org/10.1017/jfm.2018.457} {\bibfield  {journal} {\bibinfo
  {journal} {Journal of Fluid Mechanics}\ }\textbf {\bibinfo {volume} {850}},\
  \bibinfo {pages} {708} (\bibinfo {year} {2018})}\BibitemShut {NoStop}%
\bibitem [{\citenamefont {Seth}\ \emph {et~al.}(2008)\citenamefont {Seth},
  \citenamefont {Cloitre},\ and\ \citenamefont
  {Bonnecaze}}]{seth_influence_2008}%
  \BibitemOpen
  \bibfield  {author} {\bibinfo {author} {\bibfnamefont {J.~R.}\ \bibnamefont
  {Seth}}, \bibinfo {author} {\bibfnamefont {M.}~\bibnamefont {Cloitre}},\ and\
  \bibinfo {author} {\bibfnamefont {R.~T.}\ \bibnamefont {Bonnecaze}},\
  }\bibfield  {title} {\bibinfo {title} {Influence of short-range forces on
  wall-slip in microgel pastes},\ }\href@noop {} {\bibfield  {journal}
  {\bibinfo  {journal} {Journal of Rheology}\ }\textbf {\bibinfo {volume}
  {52}},\ \bibinfo {pages} {1241} (\bibinfo {year} {2008})}\BibitemShut
  {NoStop}%
\bibitem [{\citenamefont {Snoeijer}\ \emph {et~al.}(2013)\citenamefont
  {Snoeijer}, \citenamefont {Eggers},\ and\ \citenamefont
  {Venner}}]{Snoeijer2013}%
  \BibitemOpen
  \bibfield  {author} {\bibinfo {author} {\bibfnamefont {J.~H.}\ \bibnamefont
  {Snoeijer}}, \bibinfo {author} {\bibfnamefont {J.}~\bibnamefont {Eggers}},\
  and\ \bibinfo {author} {\bibfnamefont {C.~H.}\ \bibnamefont {Venner}},\
  }\bibfield  {title} {\bibinfo {title} {Similarity theory of lubricated
  {{Hertzian}} contacts},\ }\href {https://doi.org/10.1063/1.4826981}
  {\bibfield  {journal} {\bibinfo  {journal} {Physics of Fluids}\ }\textbf
  {\bibinfo {volume} {25}},\ \bibinfo {pages} {101705} (\bibinfo {year}
  {2013})}\BibitemShut {NoStop}%
\bibitem [{\citenamefont {G{\'e}raud}\ \emph {et~al.}(2013)\citenamefont
  {G{\'e}raud}, \citenamefont {Bocquet},\ and\ \citenamefont
  {Barentin}}]{geraud_confined_2013}%
  \BibitemOpen
  \bibfield  {author} {\bibinfo {author} {\bibfnamefont {B.}~\bibnamefont
  {G{\'e}raud}}, \bibinfo {author} {\bibfnamefont {L.}~\bibnamefont
  {Bocquet}},\ and\ \bibinfo {author} {\bibfnamefont {C.}~\bibnamefont
  {Barentin}},\ }\bibfield  {title} {\bibinfo {title} {Confined flows of a
  polymer microgel},\ }\href {https://doi.org/10.1140/epje/i2013-13030-3}
  {\bibfield  {journal} {\bibinfo  {journal} {European Physical Journal E}\
  }\textbf {\bibinfo {volume} {36}},\ \bibinfo {pages} {30} (\bibinfo {year}
  {2013})}\BibitemShut {NoStop}%
\bibitem [{\citenamefont {Ortega-Avila}\ \emph {et~al.}(2016)\citenamefont
  {Ortega-Avila}, \citenamefont {P{\'e}rez-Gonz{\'a}lez}, \citenamefont
  {Mar{\'\i}n-Santib{\'a}{\~n}ez}, \citenamefont {Rodr{\'\i}guez-Gonz{\'a}lez},
  \citenamefont {Aktas}, \citenamefont {Malik},\ and\ \citenamefont
  {Kalyon}}]{ortega-avila_axial_2016}%
  \BibitemOpen
  \bibfield  {author} {\bibinfo {author} {\bibfnamefont {J.~F.}\ \bibnamefont
  {Ortega-Avila}}, \bibinfo {author} {\bibfnamefont {J.}~\bibnamefont
  {P{\'e}rez-Gonz{\'a}lez}}, \bibinfo {author} {\bibfnamefont {B.~M.}\
  \bibnamefont {Mar{\'\i}n-Santib{\'a}{\~n}ez}}, \bibinfo {author}
  {\bibfnamefont {F.}~\bibnamefont {Rodr{\'\i}guez-Gonz{\'a}lez}}, \bibinfo
  {author} {\bibfnamefont {S.}~\bibnamefont {Aktas}}, \bibinfo {author}
  {\bibfnamefont {M.}~\bibnamefont {Malik}},\ and\ \bibinfo {author}
  {\bibfnamefont {D.~M.}\ \bibnamefont {Kalyon}},\ }\bibfield  {title}
  {\bibinfo {title} {Axial annular flow of a viscoplastic microgel with wall
  slip},\ }\href {https://doi.org/10.1122/1.4945820} {\bibfield  {journal}
  {\bibinfo  {journal} {Journal of Rheology}\ }\textbf {\bibinfo {volume}
  {60}},\ \bibinfo {pages} {503} (\bibinfo {year} {2016})}\BibitemShut
  {NoStop}%
\bibitem [{\citenamefont {McGraw}(2025)}]{McGraw2025}%
  \BibitemOpen
  \bibfield  {author} {\bibinfo {author} {\bibfnamefont {J.~D.}\ \bibnamefont
  {McGraw}},\ }\bibfield  {title} {{\selectlanguage {english}\bibinfo {title}
  {Transport of soft matter in complex and confined environments}},\ }\href
  {https://doi.org/10.1051/epn/2025307} {\bibfield  {journal} {\bibinfo
  {journal} {Europhysics News}\ }\textbf {\bibinfo {volume} {56}},\ \bibinfo
  {pages} {13} (\bibinfo {year} {2025})}\BibitemShut {NoStop}%
\bibitem [{Note1()}]{Note1}%
  \BibitemOpen
  \bibinfo {note} {SM contains details on: experimental protocols; fitting
  velocity profiles of Figure~\ref {fig:velocimetry}; uncertainty on the
  velocity and comparison between TIRFM and µPIV velocites; information
  supporting enhanced-viscosity lubrication layer; influence of
  solvent-microgel boundary shape on measured velocity; derivation for the
  elastohydrodynamic balance to see nanoparticles at the wall}\BibitemShut
  {NoStop}%
\bibitem [{Note2()}]{Note2}%
  \BibitemOpen
  \bibinfo {note} {The reported height corresponds to the average channel
  height under flow, as measured by confocal microscopy}\BibitemShut {NoStop}%
\bibitem [{\citenamefont {Divoux}\ \emph {et~al.}(2010)\citenamefont {Divoux},
  \citenamefont {Tamarii}, \citenamefont {Barentin},\ and\ \citenamefont
  {Manneville}}]{divoux_transient_2010}%
  \BibitemOpen
  \bibfield  {author} {\bibinfo {author} {\bibfnamefont {T.}~\bibnamefont
  {Divoux}}, \bibinfo {author} {\bibfnamefont {D.}~\bibnamefont {Tamarii}},
  \bibinfo {author} {\bibfnamefont {C.}~\bibnamefont {Barentin}},\ and\
  \bibinfo {author} {\bibfnamefont {S.}~\bibnamefont {Manneville}},\ }\bibfield
   {title} {\bibinfo {title} {Transient {Shear} {Banding} in a {Simple} {Yield}
  {Stress} {Fluid}},\ }\href {https://doi.org/10.1103/PhysRevLett.104.208301}
  {\bibfield  {journal} {\bibinfo  {journal} {Physical Review Letters}\
  }\textbf {\bibinfo {volume} {104}},\ \bibinfo {pages} {208301} (\bibinfo
  {year} {2010})}\BibitemShut {NoStop}%
\bibitem [{Note3()}]{Note3}%
  \BibitemOpen
  \bibinfo {note} {For TIRFM measurements the wall position must be determined
  \protect \emph {a priori} with a calibrating water measurement as in
  Ref.~\cite {Guyard2021}; yet, the necessity of two liquid reservoirs requires
  intermediate tubings between reservoir and chip.}\BibitemShut {Stop}%
\bibitem [{\citenamefont {Huang}\ \emph {et~al.}(2006)\citenamefont {Huang},
  \citenamefont {Guasto},\ and\ \citenamefont {Breuer}}]{Huang2006}%
  \BibitemOpen
  \bibfield  {author} {\bibinfo {author} {\bibfnamefont {P.}~\bibnamefont
  {Huang}}, \bibinfo {author} {\bibfnamefont {J.~S.}\ \bibnamefont {Guasto}},\
  and\ \bibinfo {author} {\bibfnamefont {K.~S.}\ \bibnamefont {Breuer}},\
  }\bibfield  {title} {\bibinfo {title} {Direct measurement of slip velocities
  using three-dimensional total internal reflection velocimetry},\ }\href
  {https://doi.org/10.1017/S0022112006002229} {\bibfield  {journal} {\bibinfo
  {journal} {Journal of Fluid Mechanics}\ }\textbf {\bibinfo {volume} {566}},\
  \bibinfo {pages} {447} (\bibinfo {year} {2006})}\BibitemShut {NoStop}%
\bibitem [{\citenamefont {Yoda}\ and\ \citenamefont {Kazoe}(2011)}]{Yoda2011}%
  \BibitemOpen
  \bibfield  {author} {\bibinfo {author} {\bibfnamefont {M.}~\bibnamefont
  {Yoda}}\ and\ \bibinfo {author} {\bibfnamefont {Y.}~\bibnamefont {Kazoe}},\
  }\bibfield  {title} {\bibinfo {title} {Dynamics of suspended colloidal
  particles near a wall: Implications for interfacial particle velocimetry},\
  }\href@noop {} {\bibfield  {journal} {\bibinfo  {journal} {Physics of
  Fluids}\ }\textbf {\bibinfo {volume} {23}},\ \bibinfo {pages} {111301}
  (\bibinfo {year} {2011})}\BibitemShut {NoStop}%
\bibitem [{\citenamefont {Li}\ \emph {et~al.}(2015)\citenamefont {Li},
  \citenamefont {D'eramo}, \citenamefont {Lee}, \citenamefont {Monti},
  \citenamefont {Yonger}, \citenamefont {Tabeling}, \citenamefont {Chollet},
  \citenamefont {Bresson},\ and\ \citenamefont {Tran}}]{Li2015}%
  \BibitemOpen
  \bibfield  {author} {\bibinfo {author} {\bibfnamefont {Z.}~\bibnamefont
  {Li}}, \bibinfo {author} {\bibfnamefont {L.}~\bibnamefont {D'eramo}},
  \bibinfo {author} {\bibfnamefont {C.}~\bibnamefont {Lee}}, \bibinfo {author}
  {\bibfnamefont {F.}~\bibnamefont {Monti}}, \bibinfo {author} {\bibfnamefont
  {M.}~\bibnamefont {Yonger}}, \bibinfo {author} {\bibfnamefont
  {P.}~\bibnamefont {Tabeling}}, \bibinfo {author} {\bibfnamefont
  {B.}~\bibnamefont {Chollet}}, \bibinfo {author} {\bibfnamefont
  {B.}~\bibnamefont {Bresson}},\ and\ \bibinfo {author} {\bibfnamefont
  {Y.}~\bibnamefont {Tran}},\ }\bibfield  {title} {{\selectlanguage
  {english}\bibinfo {title} {Near-wall nanovelocimetry based on total internal
  reflection fluorescence with continuous tracking}},\ }\href
  {https://doi.org/10.1017/jfm.2015.12} {\bibfield  {journal} {\bibinfo
  {journal} {Journal of Fluid Mechanics}\ }\textbf {\bibinfo {volume} {766}},\
  \bibinfo {pages} {147} (\bibinfo {year} {2015})}\BibitemShut {NoStop}%
\bibitem [{\citenamefont {Vilquin}\ \emph {et~al.}(2021)\citenamefont
  {Vilquin}, \citenamefont {Bertin}, \citenamefont {Soulard}, \citenamefont
  {Guyard}, \citenamefont {Rapha{\"e}l}, \citenamefont {Restagno},
  \citenamefont {Salez},\ and\ \citenamefont {McGraw}}]{Vilquin2021}%
  \BibitemOpen
  \bibfield  {author} {\bibinfo {author} {\bibfnamefont {A.}~\bibnamefont
  {Vilquin}}, \bibinfo {author} {\bibfnamefont {V.}~\bibnamefont {Bertin}},
  \bibinfo {author} {\bibfnamefont {P.}~\bibnamefont {Soulard}}, \bibinfo
  {author} {\bibfnamefont {G.}~\bibnamefont {Guyard}}, \bibinfo {author}
  {\bibfnamefont {E.}~\bibnamefont {Rapha{\"e}l}}, \bibinfo {author}
  {\bibfnamefont {F.}~\bibnamefont {Restagno}}, \bibinfo {author}
  {\bibfnamefont {T.}~\bibnamefont {Salez}},\ and\ \bibinfo {author}
  {\bibfnamefont {J.~D.}\ \bibnamefont {McGraw}},\ }\bibfield  {title}
  {\bibinfo {title} {Time dependence of advection-diffusion coupling for
  nanoparticle ensembles},\ }\href
  {https://doi.org/10.1103/PhysRevFluids.6.064201} {\bibfield  {journal}
  {\bibinfo  {journal} {Physical Review Fluids}\ }\textbf {\bibinfo {volume}
  {6}},\ \bibinfo {pages} {064201} (\bibinfo {year} {2021})}\BibitemShut
  {NoStop}%
\bibitem [{\citenamefont {Vilquin}\ \emph {et~al.}(2023)\citenamefont
  {Vilquin}, \citenamefont {Bertin}, \citenamefont {Rapha{\"e}l}, \citenamefont
  {Dean}, \citenamefont {Salez},\ and\ \citenamefont {McGraw}}]{Vilquin2023}%
  \BibitemOpen
  \bibfield  {author} {\bibinfo {author} {\bibfnamefont {A.}~\bibnamefont
  {Vilquin}}, \bibinfo {author} {\bibfnamefont {V.}~\bibnamefont {Bertin}},
  \bibinfo {author} {\bibfnamefont {E.}~\bibnamefont {Rapha{\"e}l}}, \bibinfo
  {author} {\bibfnamefont {D.~S.}\ \bibnamefont {Dean}}, \bibinfo {author}
  {\bibfnamefont {T.}~\bibnamefont {Salez}},\ and\ \bibinfo {author}
  {\bibfnamefont {J.~D.}\ \bibnamefont {McGraw}},\ }\bibfield  {title}
  {\bibinfo {title} {Nanoparticle {{Taylor}} dispersion near charged surfaces
  with an open boundary},\ }\href
  {https://doi.org/10.1103/PhysRevLett.130.038201} {\bibfield  {journal}
  {\bibinfo  {journal} {Physical Review Letters}\ }\textbf {\bibinfo {volume}
  {130}},\ \bibinfo {pages} {038201} (\bibinfo {year} {2023})}\BibitemShut
  {NoStop}%
\bibitem [{Note4()}]{Note4}%
  \BibitemOpen
  \bibinfo {note} {The exponent $k$ mediates the transition between linear and
  constant velocity regimes. Choosing other values of $k$ has no impact on
  general conclusions reached here but contributes to the error bars on
  $z_*$}\BibitemShut {NoStop}%
\bibitem [{\citenamefont {Mansard}\ \emph
  {et~al.}(2014{\natexlab{b}})\citenamefont {Mansard}, \citenamefont
  {Bocquet},\ and\ \citenamefont {Colin}}]{mansard_boundary_2014}%
  \BibitemOpen
  \bibfield  {author} {\bibinfo {author} {\bibfnamefont {V.}~\bibnamefont
  {Mansard}}, \bibinfo {author} {\bibfnamefont {L.}~\bibnamefont {Bocquet}},\
  and\ \bibinfo {author} {\bibfnamefont {A.}~\bibnamefont {Colin}},\ }\bibfield
   {title} {\bibinfo {title} {Boundary conditions for soft glassy flows:
  slippage and surface fluidization},\ }\href@noop {} {\bibfield  {journal}
  {\bibinfo  {journal} {Soft Matter}\ }\textbf {\bibinfo {volume} {10}},\
  \bibinfo {pages} {6984} (\bibinfo {year} {2014}{\natexlab{b}})}\BibitemShut
  {NoStop}%
\bibitem [{\citenamefont {Younes}\ \emph {et~al.}(2020)\citenamefont {Younes},
  \citenamefont {Bertola}, \citenamefont {Castelain},\ and\ \citenamefont
  {Burghelea}}]{younes_slippery_2020}%
  \BibitemOpen
  \bibfield  {author} {\bibinfo {author} {\bibfnamefont {E.}~\bibnamefont
  {Younes}}, \bibinfo {author} {\bibfnamefont {V.}~\bibnamefont {Bertola}},
  \bibinfo {author} {\bibfnamefont {C.}~\bibnamefont {Castelain}},\ and\
  \bibinfo {author} {\bibfnamefont {T.}~\bibnamefont {Burghelea}},\ }\bibfield
  {title} {{\selectlanguage {en}\bibinfo {title} {Slippery flows of a
  {Carbopol} gel in a microchannel}},\ }\href
  {https://doi.org/10.1103/PhysRevFluids.5.083303} {\bibfield  {journal}
  {\bibinfo  {journal} {Physical Review Fluids}\ }\textbf {\bibinfo {volume}
  {5}},\ \bibinfo {pages} {083303} (\bibinfo {year} {2020})}\BibitemShut
  {NoStop}%
\bibitem [{\citenamefont {Jung}\ and\ \citenamefont
  {Fielding}(2021)}]{jung_wall_2021}%
  \BibitemOpen
  \bibfield  {author} {\bibinfo {author} {\bibfnamefont {G.}~\bibnamefont
  {Jung}}\ and\ \bibinfo {author} {\bibfnamefont {S.~M.}\ \bibnamefont
  {Fielding}},\ }\bibfield  {title} {{\selectlanguage {en}\bibinfo {title}
  {Wall slip and bulk yielding in soft particle suspensions}},\ }\href
  {https://doi.org/10.1122/8.0000171} {\bibfield  {journal} {\bibinfo
  {journal} {Journal of Rheology}\ }\textbf {\bibinfo {volume} {65}},\ \bibinfo
  {pages} {199} (\bibinfo {year} {2021})}\BibitemShut {NoStop}%
\bibitem [{\citenamefont {Agnolin}\ and\ \citenamefont
  {Roux}(2007)}]{agnolin_internal_2007}%
  \BibitemOpen
  \bibfield  {author} {\bibinfo {author} {\bibfnamefont {I.}~\bibnamefont
  {Agnolin}}\ and\ \bibinfo {author} {\bibfnamefont {J.-N.}\ \bibnamefont
  {Roux}},\ }\bibfield  {title} {{\selectlanguage {en}\bibinfo {title}
  {Internal states of model isotropic granular packings. {I}. {Assembling}
  process, geometry, and contact networks}},\ }\href
  {https://doi.org/10.1103/PhysRevE.76.061302} {\bibfield  {journal} {\bibinfo
  {journal} {Physical Review E}\ }\textbf {\bibinfo {volume} {76}},\ \bibinfo
  {pages} {061302} (\bibinfo {year} {2007})}\BibitemShut {NoStop}%
\bibitem [{\citenamefont {Galvani}\ \emph {et~al.}(2023)\citenamefont
  {Galvani}, \citenamefont {Pasquet}, \citenamefont {Mukherjee}, \citenamefont
  {Requier}, \citenamefont {Cohen-Addad}, \citenamefont {Pitois}, \citenamefont
  {H{\"o}hler}, \citenamefont {Rio}, \citenamefont {Salonen}, \citenamefont
  {Durian},\ and\ \citenamefont {Langevin}}]{galvani_hierarchical_2023}%
  \BibitemOpen
  \bibfield  {author} {\bibinfo {author} {\bibfnamefont {N.}~\bibnamefont
  {Galvani}}, \bibinfo {author} {\bibfnamefont {M.}~\bibnamefont {Pasquet}},
  \bibinfo {author} {\bibfnamefont {A.}~\bibnamefont {Mukherjee}}, \bibinfo
  {author} {\bibfnamefont {A.}~\bibnamefont {Requier}}, \bibinfo {author}
  {\bibfnamefont {S.}~\bibnamefont {Cohen-Addad}}, \bibinfo {author}
  {\bibfnamefont {O.}~\bibnamefont {Pitois}}, \bibinfo {author} {\bibfnamefont
  {R.}~\bibnamefont {H{\"o}hler}}, \bibinfo {author} {\bibfnamefont
  {E.}~\bibnamefont {Rio}}, \bibinfo {author} {\bibfnamefont {A.}~\bibnamefont
  {Salonen}}, \bibinfo {author} {\bibfnamefont {D.~J.}\ \bibnamefont
  {Durian}},\ and\ \bibinfo {author} {\bibfnamefont {D.}~\bibnamefont
  {Langevin}},\ }\bibfield  {title} {\bibinfo {title} {Hierarchical bubble size
  distributions in coarsening wet liquid foams},\ }\href
  {https://doi.org/10.1073/pnas.2306551120} {\bibfield  {journal} {\bibinfo
  {journal} {Proceedings of the National Academy of Sciences}\ }\textbf
  {\bibinfo {volume} {120}},\ \bibinfo {pages} {e2306551120} (\bibinfo {year}
  {2023})},\ \bibinfo {note} {publisher: Proceedings of the National Academy of
  Sciences}\BibitemShut {NoStop}%
\bibitem [{Note5()}]{Note5}%
  \BibitemOpen
  \bibinfo {note} {In Ref.~\cite {Snoeijer2013} is provided a different
  prediction $\delta \sim R(\eta _0 v_\infty /R G_\protect \mathrm {p})^{3/5}$,
  which yields $\delta \approx 10-90$~nm.}\BibitemShut {Stop}%
\bibitem [{\citenamefont {Seth}\ \emph {et~al.}(2006)\citenamefont {Seth},
  \citenamefont {Cloitre},\ and\ \citenamefont {Bonnecaze}}]{Seth2006}%
  \BibitemOpen
  \bibfield  {author} {\bibinfo {author} {\bibfnamefont {J.~R.}\ \bibnamefont
  {Seth}}, \bibinfo {author} {\bibfnamefont {M.}~\bibnamefont {Cloitre}},\ and\
  \bibinfo {author} {\bibfnamefont {R.~T.}\ \bibnamefont {Bonnecaze}},\
  }\bibfield  {title} {\bibinfo {title} {Elastic properties of soft particle
  pastes},\ }\href {https://doi.org/10.1122/1.2186982} {\bibfield  {journal}
  {\bibinfo  {journal} {Journal of Rheology}\ }\textbf {\bibinfo {volume}
  {50}},\ \bibinfo {pages} {353} (\bibinfo {year} {2006})}\BibitemShut
  {NoStop}%
\bibitem [{\citenamefont {Zheng}\ \emph {et~al.}(2018)\citenamefont {Zheng},
  \citenamefont {Shi},\ and\ \citenamefont {Silber-Li}}]{Zheng2018}%
  \BibitemOpen
  \bibfield  {author} {\bibinfo {author} {\bibfnamefont {X.}~\bibnamefont
  {Zheng}}, \bibinfo {author} {\bibfnamefont {F.}~\bibnamefont {Shi}},\ and\
  \bibinfo {author} {\bibfnamefont {Z.}~\bibnamefont {Silber-Li}},\ }\bibfield
  {title} {{\selectlanguage {english}\bibinfo {title} {Study on the statistical
  intensity distribution ({SID}) of fluorescent nanoparticles in {TIRFM}
  measurement}},\ }\href {https://doi.org/10.1007/s10404-018-2145-2} {\bibfield
   {journal} {\bibinfo  {journal} {Microfluid Nanofluid}\ }\textbf {\bibinfo
  {volume} {22}},\ \bibinfo {pages} {127} (\bibinfo {year} {2018})}\BibitemShut
  {NoStop}%
\end{thebibliography}%

\end{document}